\documentclass[aps,preprint,showpacs,showkeys]{revtex4}

\usepackage{amsmath}
\usepackage{amssymb}
\usepackage{graphicx}
\usepackage{color}
\usepackage{verbatim}

\newcommand{\beq}{\begin{equation}}
\newcommand{\eeq}{\end{equation}}

\newcommand{\benn}{\begin{eqnarray}}
\newcommand{\eenn}{\end{eqnarray}}

\begin{document}

{\raggedleft {\it Accepted for publication on Physics of Plasmas}}
\\

\title{Gyro-induced acceleration of magnetic reconnection}


\author{L. Comisso$^{1,2,*}$, D. Grasso$^{1,2}$, F.L. Waelbroeck$^3$, D. Borgogno$^1$\\
{\small $^1$ Dipartimento Energia, Politecnico di Torino, Corso Duca degli Abruzzi 24, 10129, Torino, Italy} \\
{\small $^2$ Istituto dei Sistemi Complessi - CNR, Via dei Taurini 19, 00185, Roma, Italy} \\
{\small $^3$ Institute for Fusion Studies, The University of Texas at Austin, Austin, TX 78712-1060, USA} \\
{\small $^*$ Electronic mail: luca.comisso@polito.it}}

\baselineskip 24 pt

\begin{abstract}
The linear and nonlinear evolution of magnetic reconnection in collisionless high-temperature plasmas with a strong guide field is analyzed on the basis of a two-dimensional gyrofluid model. The linear growth rate of the reconnecting instability is compared to analytical calculations over the whole spectrum of linearly unstable wave numbers. In the strongly unstable regime (large $\Delta '$), the nonlinear evolution of the reconnecting instability is found to undergo two distinctive acceleration phases separated by a stall phase in which the instantaneous growth rate decreases. The first acceleration phase is caused by the formation of strong electric fields close to the $X$-point due to ion gyration, while the second acceleration phase is driven by the development of an open Petschek-like configuration due to both ion and electron temperature effects. Furthermore, the maximum instantaneous growth rate is found to increase dramatically over its linear value for decreasing diffusion layers. This is a consequence of the fact that the peak instantaneous growth rate becomes weakly dependent on the microscopic plasma parameters if the diffusion region thickness is sufficiently smaller than the equilibrium magnetic field scale length. When this condition is satisfied, the peak reconnection rate asymptotes to a constant value.

\end{abstract}



\maketitle

\section{Introduction}

Magnetic reconnection is a fundamental plasma process that changes the topology of the magnetic field lines and results in the conversion of magnetic energy into kinetic energy, thermal energy, and particle acceleration \cite{PriFor_2000,Bis_2000}. It is believed to be responsible for many of the most spectacular and energetic phenomena in space and laboratory plasmas. The most prominent examples include Earth magnetospheric substorms \cite{Dungey_1961}, solar and stellar flares \cite{Giova_1946}, coronal mass ejections \cite{LF_2000}, coronal heating \cite{Parker_1988}, generation of energetic particles \cite{Oier_2002}, sawtooth crashes \cite{Edw_1986} and major disruptions in tokamak experiments \cite{Boozer_2012}.

Conventional resistive magnetohydrodynamics (MHD) models are able to account for magnetic reconnection, but generally predict reconnection rates valid only for sufficiently collisional plasmas. In the well-known Sweet-Parker model of magnetic reconnection \cite{Sweet_1958,Parker_1957}, the plasma resistivity $\eta$ breaks the frozen-in flux constraint in a narrow two-dimensional bundary layer (the diffusion region) allowing magnetic field lines to reconnect. However, the elongated diffusion region distinctive of this model limits the rate of reconnection due to the Alfv\'en limit on the ion outflow velocity. In fact, assuming steady-state reconnection in an incompressible plasma, the continuity equation yields the following relation for the inflow velocity into the diffusion region
\begin{equation}
{v_{in}} \sim \frac{{{\delta_{SP}}}}{\Delta }{v_{A,up}} \ll {v_{A,up}} , \label{SP_scaling}
\end{equation}
with $\delta_{SP}$ and $\Delta$ being, respectively, the small width ($\propto \eta^{{1}/{2}}$) and the macroscopic length \cite{Waelbr_1989} of the diffusion region, and ${v_{A,up}}$ being the Alfv\'en speed based on the reconnecting component of the magnetic field just upstream of the diffusion region. Since $\delta_{SP} \ll \Delta$, the reconnection rate given in Eq. (\ref{SP_scaling}) is small and generally inconsistent with the observed fast energy release that characterizes many magnetic reconnection events \cite{Edw_1986,Oier_2001,Isobe_2005,Egedal_2007}. At small values of resistivity the development of secondary islands (plasmoids) eventually fragments the diffusion region yielding higher reconnection rates \cite{Lou_2007,Daugh_2009,BHYR_2009,SLUSC_2009,CSD_2010,SC_2010,HB_2010,ULS_2010}. In contrast, in the classical Petschek model of magnetic reconnection \cite{Petschek_1964} the outflow region forms an open (X-type) configuration, leaving a relatively short diffusion region $\Delta$ in Eq. (\ref{SP_scaling}) and, therefore, greatly enhancing the reconnection rate. However, numerical simulations showed that the open Petschek outflow geometry cannot be sustained in a model with a spatially uniform resistivity \cite{Bisk_1986}. An inhomogeneous resistivity that increases sharply in the reconnection layer facilitates a Petschek-like reconnection configuration \cite{SH_1979}, but the establishment and role of such anomalous resistivity during magnetic reconnection is not yet well understood.

In addition to the issues discussed so far, there is a further comment to be made about the reconnection rates predicted by the Sweet-Parker and Petschek models. Since these models are steady-state, they can provide only one time scale, that of steady-state reconnection, which is proportional to $S^{{1}/{2}}$ for Sweet-Parker and $\ln S$ for Petschek, where $S = {{{\mu_0}{\Delta}{v_{A,up}}}/{\eta}}$ is the Lundquist number and $\mu_0$ is the vacuum permeability. In contrast, reconnection in nature is generally not a steady-state process, but rather a dynamical one. In particular, there are many magnetic reconnection phenomena in laboratory as well as space and astrophysical plasmas where the dynamics exhibits an impulsive behaviour, i.e. a sudden increase in the time derivative of the reconnection rate \cite{Bhatta_2004,Egedal_2007,Yam_2011}. This is often referred to as the ``onset problem'', which addresses why the magnetic field configuration evolves slowly for a long period of time, only to undergo an abrupt dynamical change over a much shorter period of time \cite{Waelbr_1993,CSD_2005,HBS_2011}. It is therefore necessary to move beyond the steady-state models in order to explain the dynamics of fast magnetic reconnection phenomena. A significant step forward, aimed at understanding fast sawtooth crashes in tokamaks, was obtained when Aydemir showed, by means of numerical simulations, that in strongly unstable semicollisional/collisionless regimes a relatively slow initial phase of the reconnection process is followed by a dramatic acceleration caused by electron pressure gradients \cite{A_1992}. Aydemir's results were corroborated one year later by Wang and Bhattacharjee \cite{WB_1993}, while Ottaviani and Porcelli \cite{OP_1993} showed that electron inertia, by itself, can lead to growth rates faster-than-exponential in time. It is important to note that in these works the nonlocal ion response was neglected since it was believed that the two-fluid theory was adequate to properly describe the reconnection dynamics \cite{ZR_1992}. In the present paper we show that including the correct gyrofluid response does make a difference. In particular, we have found that more than one nonlinear acceleration is possible when ion gyration effects are taken into account. We will give numerical and analytical evidence that the qualitative difference between hot and cold ion reconnection is linked to the formation of strong electric fields due to ion gyration effects. Furthermore, we will discuss how the microscopic plasma parameters affect both the slow initial phase and the fast nonlinear phase of the reconnection process.

We are interested in the regime of rarefied high-temperature plasmas in which the collisional mean free path is large enough that classical Coulomb collisions are negligible. 
Because of the relevance in many cases of physical interest, we consider magnetic reconnection phenomena that take place in a two-dimensional plane perpendicular to a strong and essentially uniform component of the magnetic field, the so-called ``guide field''. The presence of this strong background magnetic field creates a spatial anisotropy that makes it possible to exploit the ordering ${k_\parallel} \ll {k_\bot}$, where ${k_\parallel}$ and ${k_\bot}$ are the typical wave numbers of the fluctuation spectrum in the direction parallel and perpendicular to the equilibrium magnetic field. The reconnecting component of the magnetic field is small compared to the total magnetic field strength. More generally the amplitude of the fluctuating fields is assumed to be small, while their perpendicular gradients can be comparable to or larger than those of the equilibrium fields. Moreover, the strong guide field ensures that the time variations associated with reconnection are slow compared to the ion gyro-period. These features are necessary to adopt a gyrofluid approach to the study of magnetic reconnection. The gyrofluid choice allows us to investigate ion and ion-sound Larmor radius effects (that cannot be neglected in high-temperature plasmas) within the framework of a generalized fluid model, which is computationally less expensive and physically more intuitive than a kinetic one. 
The model equations are presented in the next section, while in the subsequent sections this gyrofluid model is used to study magnetic reconnection in a current sheet. Finally, the most relevant results are summarized in the concluding section.

\section{Model Equations}  \label{sec2}

As discussed in the previous section, we are interested in a model that can describe two-dimensional magnetic reconnection phenomena in collisionless high-temperature plasmas embedded in a strong and uniform magnetic field. For this purpose we consider an isothermal gyrofluid model that can be obtained from the equations of Ref. \cite{WT_2012} by neglecting magnetic curvature effects and assuming that all the fields are translationally invariant along the direction of the strong guide field ${B_0}{\bf{\hat z}}$, which is perpendicular to the reconnection plane. The pressure is assumed to be scalar for both the electrons and the ions, and the electron inertia provides the mechanism for breaking the frozen-in flux constraint. A right-handed Cartesian coordinate system $(x,y,z)$ is adopted, and a plasma with single ion species and charge number $Z=1$ is assumed.

We adopt a normalization scheme such that all the lengths are normalized to a characteristic equilibrium magnetic field scale length $L$, and all times to the Alfv\'en time $\tau_A={L}/{v_A}$, where $v_A={B_0}/{{\left( {{\mu_0}{n_0}{m_i}} \right)^{{1}/{2}}}}$, with $n_0$ a constant background density and $m_i$ the ion mass. 
Thus, dependent variables are normalized in the following way:
\begin{eqnarray} \label{modsp}
\left( {{{\hat n}_i},{{\hat n}_e},{{\hat u}_i},{{\hat u}_e},\hat \psi ,\hat \phi } \right) = \left( {\frac{L}{{{d_i}}}\frac{{{n_i}}}{{{n_0}}},\frac{L}{{{d_i}}}\frac{{{n_e}}}{{{n_0}}},\frac{L}{{{d_i}}}\frac{{{u_i}}}{{{v_A}}},\frac{L}{{{d_i}}}\frac{{{u_e}}}{{{v_A}}},\frac{\psi }{{{B_0}L}},\frac{\phi }{{{B_0}L{v_A}}}} \right),
\end{eqnarray}
where dimensionless quantities appear on the left hand side. Hereafter the carets denoting normalized quantities will be omitted for simplicity of notation. The fields $n_i$ and ${u_i} = {\bf{\hat z}} \cdot {{\bf{\bar v}}_i}$ represent the perturbed density and the out-of-plane velocity of the ion {\it guiding centers}, whereas $n_e$ and ${u_e} = {\bf{\hat z}} \cdot {{\bf{v}}_e}$ are the perturbed density and the out-of-plane velocity of the electrons. We indicate with $\psi  = {\bf{\hat z}} \cdot {\bf{A}}$ the in-plane magnetic flux function of a magnetic field
\begin{equation}
{\bf{B}} = {\bf{\hat z}} + \nabla \psi  \times {\bf{\hat z}} ,
\end{equation}
where ${\bf{A}}$ is a vector potential. The electrostatic potential is denoted by $\phi$, hence the electric field can be expressed as
\begin{equation}
{\bf{E}} = - \frac{{\partial \psi }}{{\partial t}} {\bf{\hat z}} - \nabla \phi .
\end{equation}

The evolution equations of our model consist of the continuity equation and the $z$-component of the equation of motion for the ion guiding centers:
\begin{equation}
\frac{\partial n_i}{\partial t}+[\Phi,n_i] = [\Psi, u_i] ,\label{e1}
\end{equation}
\begin{equation}
\frac{\partial }{{\partial t}}\left( {\Psi + d_i^2{u_i}} \right)+[\Phi,\Psi+d_i^2 u_i] = \rho_i^2 [\Psi, n_i] ,\label{e2}
\end{equation}
and similar equations for the electrons, where the vanishingly small electron Larmor radius limit ${\rho_e} \to 0$ is taken:
\begin{equation}
\frac{\partial n_e}{\partial t}+[\phi,n_e] = [\psi, u_e] ,\label{e3}
\end{equation}
\begin{equation}
\frac{\partial }{{\partial t}}\left( {\psi - d_e^2{u_e}} \right)+[\phi,\psi-d_e^2 u_e] = -\rho_s^2 [\psi, n_e] .
\label{e4}
\end{equation}
The symbol $\left[ { \cdot , \cdot } \right]$ denotes the canonical Poisson bracket, so that $\left[ {f,g} \right] \equiv {\bf{\hat z}} \cdot \left( {\nabla f \times \nabla g} \right)$ for two generic fields $f$ and $g$, whereas the four dimensionless parameters appearing in the above equations are the (normalized) electron and ion skin depth, $d_e={\left({{c}/{\omega_{pe}}}\right)}/{L}$ and $d_i={\left({{c}/{\omega_{pi}}}\right)}/{L}$ respectively, and the (normalized) ion and ion-sound Larmor radius, $\rho_i={\left({{v_{ti}}/{\omega_{ci}}}\right)}/{L}$ and $\rho_s={\left({{c_{se}}/{\omega_{ci}}}\right)}/{L}$ respectively. Here, $c_{se}={\left( {{T_e}/{m_i}} \right)^{{1}/{2}}}$ is the sound speed based on the electron temperature, $v_{ti}={\left( {{T_i}/{m_i}} \right)^{{1}/{2}}}$ is the ion thermal speed, and the other symbols have their usual meaning. Furthermore, 
\begin{equation}
\Phi=\Gamma_0^{{1}/{2}} \phi, \qquad \Psi=\Gamma_0^{{1}/{2}} \psi, \label{gyro_relations}
\end{equation}
are the gyro-averaged $\phi$ and $\psi$, where the symbol $\Gamma_0^{{1}/{2}}$ refers to the gyro-averaged operator introduced by Dorland and Hammet \cite{DH_1993} that we adopt in its lowest-order Pad\'e approximant form 
\begin{equation}
\Gamma_0^{{1}/{2}} = \dfrac{1}{{1 - \dfrac{{\rho_i^2}}{2}\nabla_\bot^2}} ,
\end{equation}
which is valid for arbitrary $k_\bot ^2 \rho_i ^2$. Note that the ion guiding centers do not respond to the local electromagnetic field but to the field averaged over its gyro-orbit. Therefore the ion guiding centers are advected by their nonlocal value of the electric drift, related to the gyro-averaged electrostatic potential according to ${\bf{\bar v}}_E = {\bf{\hat z}} \times \nabla \Phi$. Since the present model neglects the electron Larmor gyration, the electrons are instead advected by their local value of the electric drift ${\bf{v}}_E = {\bf{\hat z}} \times \nabla \phi$.

Eqs. (\ref{e1})-(\ref{e4}) are closed by the $z$-component of Amp\`ere's law
\begin{equation}
\nabla _ \bot ^2\psi  =  - j =  - \Gamma_0^{1/2}{u_i} + {u_e}, \label{e5}
\end{equation}
where $j = {\bf{\hat z}} \cdot {\bf{J}}$ is the out-of-plane current density, and by imposing quasi-neutrality on the particle density (not the guiding-center density)
\begin{equation}
n_e=\Gamma_0^{1/2}n_i +\left(\frac{\Gamma_0-1}{\rho_i^2}\right)\phi, \label{e6}
\end{equation}
with $\Gamma_0=(\Gamma_0^{{1}/{2}})^2$. In the above equation, the term $\Gamma_0^{1/2}n_i$ is the gyrophase-independent part of the real space ion particle density, whereas the term ${{\left( {{\Gamma_0} - 1} \right)\phi}/{\rho_i^2}}$, which arises from the gyrophase-dependent part of the distribution function, represents the polarization density due to the variation of the electric field around a gyro-orbit.

The evolution equations of the model conserve the following energy integral:
\begin{equation}
H = \frac{1}{2}\int_{\mathcal{D}} {{d^2}x(|\nabla \psi{|^2} + d_i^2u_i^2 + d_e^2u_e^2 + \rho_i^2n_i^2 + \rho_s^2n_e^2 + \Phi {n_i} - \phi {n_e})}, \\ \label{ham}
\end{equation}
where we have used Amp\`ere's law and the quasi-neutrality equation to simplify the result. Here, $\mathcal{D}$ denotes the spatial domain of interest, and the boundary conditions have been assumed to be such that the surface integrals vanish. The successive terms in the functional (\ref{ham}) represent, respectively, the magnetic energy, the $z$-component of the ion and electron kinetic energies, the ion and electron thermal energies, and the electrostatic energy of the ions and electrons. 
Taking the energy functional as the Hamiltonian of our 4-field model, the set of Eqs. (\ref{e1})-(\ref{e4}) can be cast into noncanonical Hamiltonian form 
\begin{equation}
\frac{{\partial {\chi^i}}}{{\partial t}} = \left\{ {{\chi^i},H} \right\},\quad i = 1,...,4  ,
\end{equation}
where $\chi^i$ are suitable field variables and $\left\{ { \cdot , \cdot } \right\}$ is the noncanonical Poisson bracket consisting of a bilinear, antisymmetric form satisfying the Leibniz rule and the Jacobi identity. 
Adopting $n_i$, $D \equiv \Psi + d_i^2{u_i}$, $n_e$, and $F \equiv \psi - d_e^2{u_e}$ as field variables, i.e. $\chi  \equiv \left( {{n_i},D,{n_e},F} \right)$, the noncanonical Poisson bracket found in Ref. \cite{WT_2012} in the limit of no magnetic curvature and ${\partial  \mathord{\left/
 {\vphantom {\partial  {\partial z}}} \right.
 \kern-\nulldelimiterspace} {\partial z}} = 0$ reduces to
\begin{equation}
\begin{split}
\left\{ {C,G} \right\} & = \int_{\mathcal{D}} {{d^2}x \left( {{n_i}\left( { - \left[ {{C_{{n_i}}},{G_{{n_i}}}} \right] - \rho_i^2 d_i^2\left[ {{C_D},{G_D}} \right]} \right) + {n_e}\left( {\left[ {{C_{{n_e}}},{G_{{n_e}}}} \right] + \rho_s^2 d_e^2\left[ {{C_F},{G_F}} \right]} \right)} \right.} \label{brpoi}\\
& \quad \left. { + D\left( { - \left[ {{C_{{n_i}}},{G_D}} \right] - \left[ {{C_D},{G_{{n_i}}}} \right]} \right) + F\left( {\left[ {{C_F},{G_{{n_e}}}} \right] - \left[ {{C_{{n_e}}},{G_F}} \right]} \right)} \right)
\end{split}
\end{equation}
for two generic functionals $C$ and $G$, with subscripts indicating functional derivatives. Noncanonical Poisson brackets are characterized by the presence of Casimir invariants (see, e.g., Ref. \cite{Morr_1998}), which are defined as non-zero functionals $C$ of the field variables that satisfy the relation $\left\{ {F,C} \right\} = 0$ for any functional $F$ of the field variables. Given that, in particular, they commute with any Hamiltonian functional, Casimir invariants are constants of motion for the system. In the case of the bracket (\ref{brpoi}), the following four infinite families of Casimirs invariants can be obtained:
\begin{equation}
\begin{split}
{C_1} = \int_{\mathcal{D}} {{d^2}}x {f_ + }\left( {D + {d_i}{\rho _i}{n_i}} \right),\\
{C_2} = \int_{\mathcal{D}} {{d^2}}x {f_ - }\left( {D - {d_i}{\rho _i}{n_i}} \right),\\
{C_3} = \int_{\mathcal{D}} {{d^2}}x {g_ + }\left( {F + {d_e}{\rho _s}{n_e}} \right),\\
{C_4} = \int_{\mathcal{D}} {{d^2}}x {g_ - }\left( {F - {d_e}{\rho _s}{n_e}} \right), \label{casimirs}
\end{split}
\end{equation}
where $f_{\pm}$ and $g_{\pm}$ represent arbitrary functions of their arguments. The form of the Casimirs (\ref{casimirs}) suggests the introduction of a new set of variables,
 \begin{eqnarray}  
 I_{\pm} \equiv D\pm d_i\rho_in_i , \qquad G_{\pm} \equiv F\pm d_e \rho_s n_e , \label{nfi}  
 \end{eqnarray} 
in terms of which Eqs. (\ref{e1})-(\ref{e4}) can be rewritten in the following form of advection equations:
\begin{equation}
\frac{{\partial {I_ \pm }}}{{\partial t}} + [{\Phi _ \pm },{I_ \pm }] = 0,\quad \frac{{\partial {G_ \pm }}}{{\partial t}} + [{\phi _ \pm },{G_ \pm }] = 0, \label{lagr_eqs}
\end{equation}
where
\begin{eqnarray}
 \Phi_{\pm} \equiv \Phi\mp\frac{\rho_i}{d_i} \Psi, \qquad \phi_{\pm} \equiv \phi\pm\frac{\rho_s}{d_e} \psi ,           \label{stre}
 \end{eqnarray} 
are the stream functions of the velocity fields ${{{\bf{\bar v}}}_\pm} = {\bf{\hat z}} \times \nabla {\Phi_\pm}$ and ${{\bf{v}}_\pm } = {\bf{\hat z}} \times \nabla {\phi_\pm}$, that advect the fields $I_{\pm}$ and $G_{\pm}$, respectively. The form of Eqs. (\ref{lagr_eqs}) make it clear that the conserved fields associated with the Casimirs preserve their initial topology. In Ref. \cite{CGTW_2012} it was shown that the investigation of these Lagrangian invariant fields helps to understand how the reconnection evolution is affected by the plasma $\beta$ and by the ratio of species temperatures.

The set of gyrofluid equations presented in this section describe the low-frequency dynamics ($\omega \ll {\omega_{ci}, k{v_A}}$) of low-$\beta$ plasmas (${\beta} \ll 1$) in the presence of a strong guide field (${B_0} \gg {B_\bot}$), and thus by assumption exclude whistler and compressional Alfv\'en waves. Here the total plasma beta is $\beta = \beta_e + \beta_i$, where the electron and ion beta are defined as ${\beta _{e,i}} \equiv {2{\mu _0}{n_0}{T_{e,i}}}/{B_0^2}$. Both the inertial ($\beta_e \ll 2{m_e}/m_i$) and the kinetic ($\beta_e \gg 2{m_e}/m_i$) Alfv\'en wave regimes are described (see Appendix), whereas for ${\beta_e} \sim 2{{m_e}/{m_i}}$ (corresponding to $v_{te} \sim v_A$) the model equations need to be extended to account for the electron Landau damping \cite{HDP_1992}. Since resistivity is neglected, the validity of the model requires also that the time scales of interest are shorter than the electron-ion collision time ($\omega \gg \nu_{ei}$). 

\section{Equilibrium Configuration}  \label{sec3}

In order to investigate the evolution of magnetic reconnection instabilities, the system of Eqs. (\ref{e1})-(\ref{e6}) is solved numerically considering an equilibrium which is linearly unstable with respect to tearing (or ``reconnecting'') modes, which tear and reconnect the magnetic field at their associated resonant surfaces defined by ${\bf{k}} \cdot {{\bf{B}}_{eq}} = 0$, where $\bf{k}$ is the wave vector of the mode and ${{\bf{B}}_{eq}}$ is the equilibrium magnetic field. In particular, we adopt the following static equilibrium:
\begin{equation}
\qquad n_{i,eq}(x)= n_{e,eq}(x)= n_{eq}, \qquad u_{i,eq}(x)=0, \qquad \psi_{eq}(x)= \sum_{n=-11}^{11} \hat f_n {e^{inx}}, \label{equil}
\end{equation}
where $n_{eq}$ represents a uniform, nondrifting background density, and $\hat f_n$ are the Fourier coefficients of the function
\begin{equation}
f\left( x \right) = \frac{{{A_0}}}{{{{\cosh }^2}\left( {\dfrac{x}{L}} \right)}}
\end{equation}
with $L = 1$ and $A_0$ representing a parameter that determines the strength of the in-plane equilibrium magnetic field. In the following we consider $A_0 = 0.1$, so that ${\max \left| {{B_{y,eq}}} \right|}/B_0 \approx 0.08$. Moreover, if we define the equilibrium magnetic shear length as $L_s = {B_0}/{{\left( {{d{B_{y,eq}}}/{dx}} \right)}}$ evaluated at the resonant surface $x=0$, choosing $L = 1$ and $A_0 = 0.1$ implies $L_s = 5$.

The fields of the model are decomposed in a time independent equilibrium and an evolving perturbation that is advanced in time according to a third order Adams-Bashforth algorithm. 
Double periodic boundary conditions are imposed and a pseudospectral method is used in a domain $\{(x,y):-\pi\leq x < \pi, -a\pi\leq y < a\pi\}$, with a resolution up to 4096$\times$512 grid points. Numerical filters are introduced acting only on typical length scales much smaller than any other physical length scale of the system. These filters smooth out the small spatial scales below a chosen cutoff, while leaving unchanged the large scale dynamics even on long times, as described in Ref. \cite{Lele_1992}.

\begin{figure}[h!]
\begin{center}
\includegraphics[bb = 10 10 298 285, width=8.6cm]{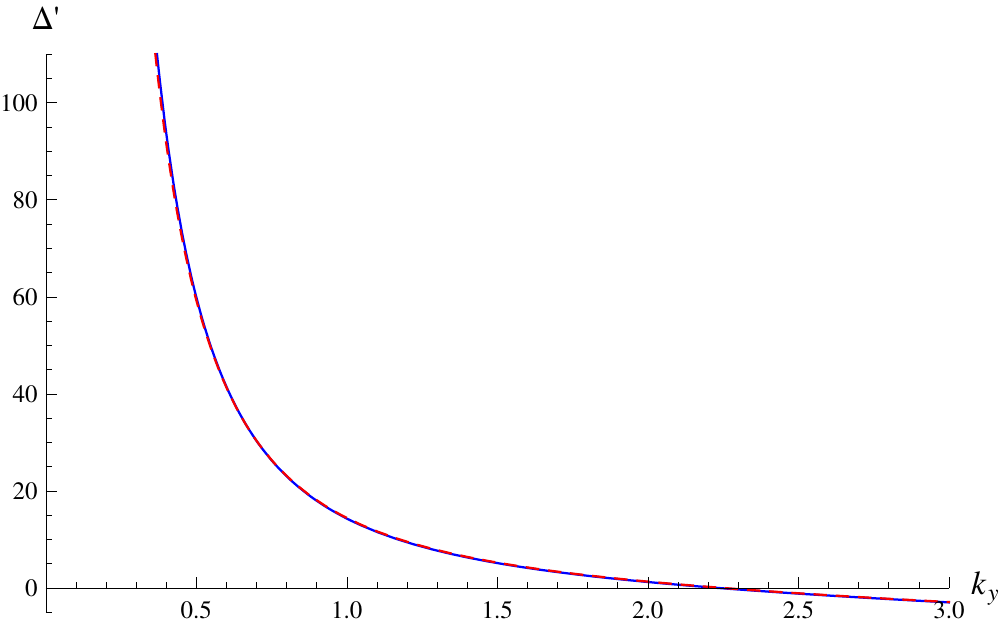}
\end{center}
\caption{(Color online) Linear tearing stability index $\Delta'$ as a function of the wave number $k_y$. The blue solid line refers to the analytical expression (\ref{Deltap_ky}), whereas the red dashed line corresponds to the numerical solution with a domain $-\pi \leq x \leq \pi$.}
\label{fig1}
\end{figure}

Note that the parameter $a$ fixes the domain length along the $y$-direction, $L_y$, which in turn is linked to the linear tearing stability index $\Delta '$ of our equilibrium. Indeed, for $\psi_{eq}(x)={A_0}/\cosh^2 x$ the following analytic form for $\Delta '$ can be obtained \cite{Porc_2002}:
\begin{equation}
\Delta ' \equiv \mathop {\lim }\limits_{\epsilon  \to 0} \left( {{{\left. {\frac{{d\ln {\psi _L}}}{{dx}}} \right|}_{ + \epsilon }} - {{\left. {\frac{{d\ln {\psi _L}}}{{dx}}} \right|}_{ - \epsilon }}} \right)
= 2 \frac{{\left( {3 + k_y^2} \right)\left( {5 - k_y^2} \right)}}{{k_y^2\sqrt {4 + k_y^2} }}, \label{Deltap_ky}
\end{equation}
where $\psi_L$ is the ideal MHD magnetic flux eigenfunction, $\epsilon$ denotes the distance from the resonant surface located at $x=0$, and $k_y = {{2\pi m}/{L_y}}$ is the wave number, with $m$ positive integer. Modes are destabilized if $\Delta ' > 0$ \cite{FKR_1963}, i.e. when ${k_y} < \sqrt 5$ for our equilibrium. Domain boundary effects can lead to a modification of the expression for the tearing stability index, however, for the equilibrium (\ref{equil}) our choice of the domain size in the $x$-direction is sufficient to avoid these effects, as shown in Fig. \ref{fig1}, where the curve of the analytical expression (\ref{Deltap_ky}) (blue solid line) is almost indistinguishable from that of the numerical solution with a domain $-\pi \leq x \leq \pi$ (red dashed line). 
 
The reconnecting instability is initiated by perturbing the equilibrium with a small disturbance on the out-of-plane current density of the form $\delta j\left( {x,y} \right) = \delta j\left( x \right)\cos ({{2\pi y} \mathord{\left/
 {\vphantom {{2\pi y} {{L_y}}}} \right.
 \kern-\nulldelimiterspace} {{L_y}}})$, where $\delta j\left( x \right)$ is a function localized within a width of the order $d_e$ around the rational surface $x=0$.

\section{Linear Phase} \label{sec4}

The linear phase of the reconnecting instability is investigated by comparing the gyrofluid growth rates with analytical calculations over a range of parameters such as $\Delta'$, ${{T_i}/{T_e}}$ and $\beta$, with and without taking into account ion acoustic waves. We focus on high-temperature plasmas characterized by ${\rho_\tau^2} \gg {d_e^2}$, i.e. $\beta \gg 2{{m_e}/{m_i}}$, where
\begin{equation}
\rho _\tau = \frac{{{{{c_s}/{\omega_{ci}}}}}}{{{L}}} = \rho_s \left( {1 + \frac{{{T_i}}}{{{T_e}}}} \right)^{{1}/{2}} = d_i {\left( \frac{\beta }{2} \right)^{{1}/{2}}} ,
\end{equation}
with $c_s = {\left( {{(T_e + T_i)}/{m_i}} \right)^{{1}/{2}}}$ being the sound speed based on both the electron and ion temperatures. In this regime, the linear dispersion relation of collisionless tearing modes was obtained analytically in Ref. \cite{P_1991} by adopting boundary layer and asymptotic matching techniques. The dispersion relation derived in this work is based on a two-fluid model in which electrons are assumed to be isothermal within the tearing layer. This is a valid approximation if $\gamma_L^2 \ll k_\parallel^2 v_{te}^2$, where $\gamma_L$ is the linear growth rate of the mode, $v_{te}={\left( {{T_e}/{m_e}} \right)^{{1}/{2}}}$ is the electron thermal velocity, and ${k_\parallel }\left( x \right) = {k_y}{{{B_{y,eq}}\left( x \right)}/{B_{eq}}} \approx {k_y}{{x}/{L_s}}$ within the tearing layer. Finite ion Larmor radius effects have been included by adopting a Pad\'e approximation of the ion response which is valid for arbitrary $k_\bot \rho_i$, while the ion acoustic wave dynamics have been ignored by assuming $\gamma _L^2 \gg k_\parallel ^2 c_s^2$. In particular, it was shown that as long as diamagnetic effects can be neglected under the assumption $\gamma_L \gg \omega _{ * e,i}$, where ${\omega _{ * e,i}}$ are electron/ion diamagnetic drift frequencies, the dispersion relation of the collisionless tearing mode in the relevant limit ${\rho_\tau} > {d_e}$ and $\hat \gamma < {\rho_\tau}$ is
\begin{equation}
\frac{\pi }{2} {\hat \gamma ^2} = {\rho_\tau }{\lambda_H} + \frac{{{\rho_\tau ^2{d_e}}}}{{{\hat \gamma}}} , \label{Porcelli_DR1}
\end{equation}
where $\hat \gamma = {{\gamma _L}/{({k_y} {B}_{y,eq}^\prime)}}$, with ${{B}_{y,eq}^\prime} = {{d{B_{y,eq}}}/{dx}}$ evaluated at the resonant surface. The parameter $\lambda_H$ is a measure of the potential energy that is available outside the tearing layer, and is linked to the linear tearing stability index by the relation ${\lambda_H} = {{- \pi}/{\Delta '}}$. Therefore, Eq. (\ref{Porcelli_DR1}) can be rewritten as
\begin{equation}
\frac{{\gamma _L^3}}{{k_y^3 {B_{y,eq}^{\prime 3}}}} = \frac{2}{\pi }{d_e}\rho _\tau ^2\left( {1 - \frac{{{\gamma _L}}}{{{k_y {{B}_{y,eq}^\prime}}}}\frac{\pi }{{{\rho _\tau }{d_e}\Delta '}}} \right) . \label{Porcelli_DR2}
\end{equation}
In the limit $\Delta ' {\rho_\tau}^{1/3} {d_e}^{2/3} \gg 1$, or more conservatively $d_e \Delta ' \gg 1$, the above dispersion relation reduces to
\begin{equation}
{\gamma_L} = {k_y}{{B_{y,eq}^\prime}} {\left( {\frac{2}{\pi }} \right)^{1/3}} {d_e}^{1/3} {\rho_\tau}^{2/3} ,
\label{DR2_high_Deltap}
\end{equation}
while in the limit $\Delta ' {\rho_\tau}^{1/3} {d_e}^{2/3} \ll 1$, neglecting the left-hand side of Eq. (\ref{Porcelli_DR2}), we obtain
\begin{equation}
{\gamma_L} = {k_y}\frac{{B_{y,eq}^\prime}}{\pi}{d_e}{\rho_\tau}\Delta' .
\label{DR2_small_Deltap}
\end{equation}
Note that for ${\rho_\tau ^2} \ll {d_e^2}$, equivalent to $\beta \ll 2{{m_e}/{m_i}}$, the ion effects are negligible and the electron response within the tearing layer is expected to be adiabatic with $\gamma_L^2 \gg k_\parallel^2 v_{te}^2$ for $\left| x \right| \lesssim {d_e}$. Hence Eq. (\ref{Porcelli_DR2}) is not valid anymore, and the dispersion relation becomes \cite{BC_1981} $\gamma_L = {k_y} {B_{y,eq}^\prime} {d_e}$ in the limit $d_e \Delta ' \gg 1$, while it yields \cite{Coppi_1964} $\gamma_L = 0.22 {k_y} {B_{y,eq}^\prime} {d_e}^3 {\Delta'}^2$ in the limit $d_e \Delta ' \ll 1$.

\begin{figure}[h!]
\begin{center}
\includegraphics[bb = 10 10 298 185, width=8.6cm]{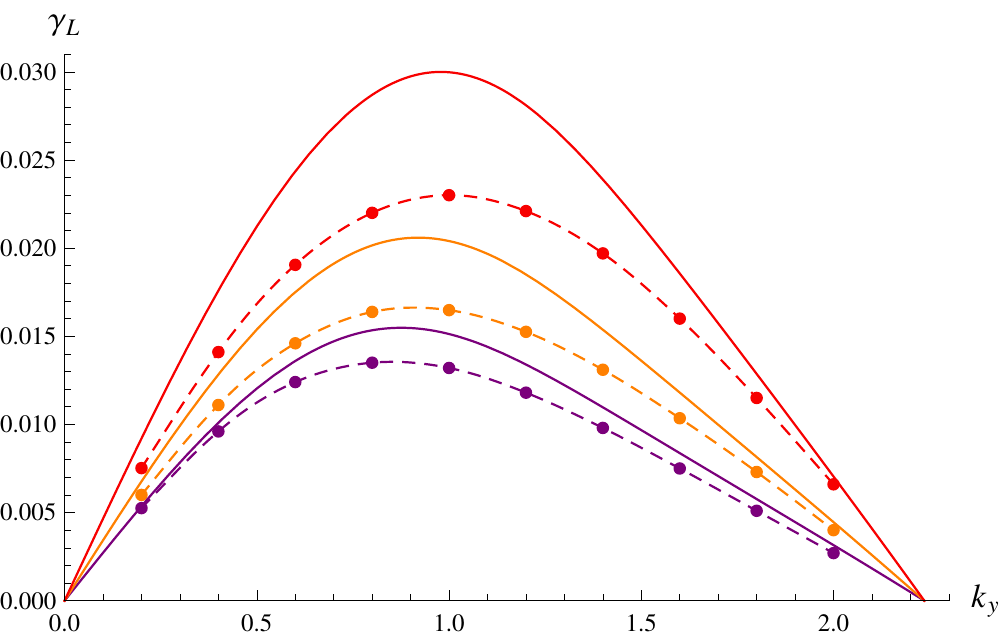}
\end{center}
\caption{(Color online) Linear growth rate $\gamma_L$ as a function of the wave number $k_y$ for the equilibrium specified in Sec. \ref{sec3} and the following plasma parameters: $d_e = 0.1$, $d_i = 1$, $\rho_s = 0.2$, and $\rho_i = {\left( {{T_i}/{T_e}} \right)^{{1}/{2}}} \rho_s = \left\{ {10^{-5},0.2,0.4} \right\}$. Different colors refer to ${{T_i}/{T_e}} \ll 1$ (purple, bottom two lines), ${{T_i}/{T_e}} = 1$ (orange, middle two lines), and ${{T_i}/{T_e}} = 4$ (red, upper two lines). The dots are values obtained from the numerical solution of the gyrofluid model, whereas the solid lines are the solution of Eq. (\ref{Porcelli_DR2}).}
\label{fig2}
\end{figure}

\begin{figure}[h!]
\begin{center}
\includegraphics[bb = 10 10 298 185, width=8.6cm]{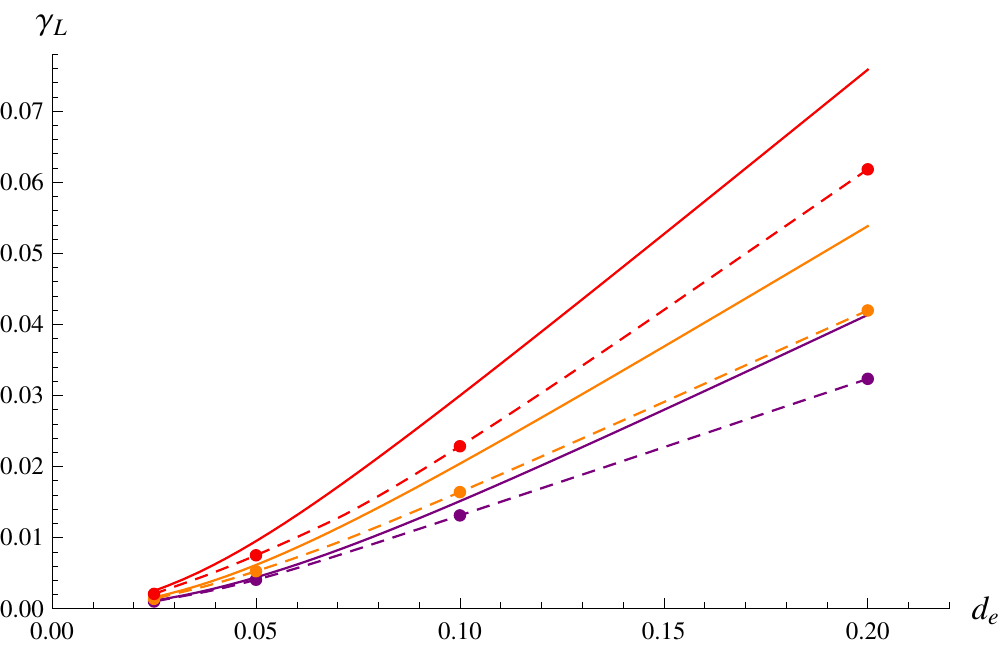}
\end{center}
\caption{(Color online) Linear growth rate $\gamma_L$ as a function of the electron skin depth $d_e$ for the wave number $k_y=1$. Plasma parameters are such that ${{d_i}/{d_e}} = {\left( {{m_i}/{m_e}} \right)^{{1}/{2}}} = 10$, ${{\rho_s}/{d_i}} = {\left( {{\beta_e}/{2}} \right)^{{1}/{2}}} = 0.2$, and ${{\rho_i}/{d_i}} = {\left( {{\beta_i}/{2}} \right)^{{1}/{2}}} = \left\{ {10^{-5},0.2,0.4} \right\}$. The equilibrium configuration, as well as the notation, is the same as in Fig. \ref{fig2}.}
\label{fig3}
\end{figure}

\begin{figure}[h!]
\begin{center}
\includegraphics[bb = 10 10 298 185, width=8.6cm]{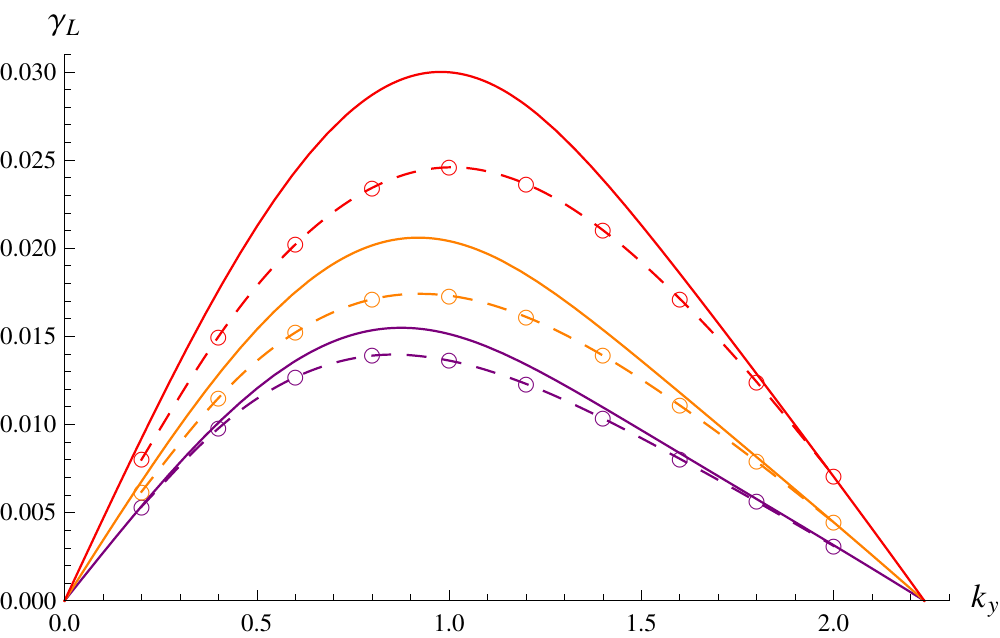}
\end{center}
\caption{(Color online) Linear growth rate $\gamma_L$ as a function of the wave number $k_y$ for $d_i = 10^{6}$ (this choice identifies the case $d_i \to \infty$, which has the effect of removing ion acoustic waves). Other plasma parameters and the equilibrium configuration are the same as in Fig. \ref{fig2}. Even the notation is the same, except that the values obtained from gyrofluid simulations are here denoted by empty circles.}
\label{fig4}
\end{figure}

Recently, two careful studies \cite{Rogers_2007,Pueschel_2011} have compared the linear growth rates obtained from gyrokinetic simulations to the analytic dispersion relation of collisionless tearing modes in the large $\Delta '$ (small $k_y$) and small $\Delta '$ (large $k_y$) regimes. Here we consider also the intermediate $\Delta '$ regime by numerically solving the complete dispersion relation, Eq. (\ref{Porcelli_DR2}), and comparing it to the gyrofluid growth rates over the whole linearly unstable $k_y$-spectra. The analytic results are shown as solid lines in Figs. \ref{fig2}-\ref{fig4}, while the results obtained from gyrofluid simulations are interpolated with dashed lines. Different ${{T_i}/{T_e}}$ ratios are considered in order to evaluate the temperature dependence of the linear growth rate. Fig. \ref{fig2} refers to the following plasma parameters: $d_e=0.1$, $d_i=1$, $\rho_s=0.1$, $\rho_i= \left\{ {10^{-5},0.2,0.4} \right\}$. The choice of the electron and ion skin depth has led to an artificial electron to ion mass ratio, ${{m_e}/{m_i}} = {\left( {{d_e}/{d_i}} \right)^{2}} = {10^{-2}}$, but on the other hand has allowed to reduce the computational resources. Since $\rho_s = d_i \left({{\beta_e}/{2}}\right)^{1/2}$, then $\beta_e$ and $T_e$ are held fixed in all cases, while the scanning of $\rho_i=\rho_s \left({{T_i}/{T_e}}\right)^{1/2}$ has the effect of varying $T_i$ as well as $\beta_i$. Fig. \ref{fig2} shows that tearing modes are unstable for ${k_y} < \sqrt 5$, i.e. for $\Delta ' > 0$, and their growth rate increases with the ion temperature. However, the growth rates obtained from gyrofluid simulations have a weaker dependence on ${{T_i}/{T_e}}$ and $\beta_i$ than the analytic theory. An even lower sensitivity to ${{T_i}/{T_e}}$ and $\beta_i$ was found in the gyrokinetic calculations of Refs. \cite{Rogers_2007} and \cite{Pueschel_2011}, where a good agreement with the relations (\ref{DR2_high_Deltap}) and (\ref{DR2_small_Deltap}) was found for low-$\beta$ plasmas, whereas for the cases with $\beta \sim 1$ and $T_i \gtrsim T_e$ the analytic theory cannot confirm their results since its validity requires $\beta < 2 \left({{m_e}/{m_i}}\right)^{1/4}$. In the low-$\beta$ regimes considered here, we find a very close agreement with the analytic dispersion relation in the small and large $\Delta '$ regimes, corresponding to the extreme right and left regions of Fig. \ref{fig2} respectively. The apparent discrepancy in the intermediate $\Delta '$ regime is resolved for lower $d_e$ (and $\rho_\tau$) values, as shown in Fig. \ref{fig3}. This is due to the fact that the analytic theory is based on asymptotic matching techniques that are increasingly accurate for thinner width layers.

The maximum linear growth rate $\gamma_{L,\max}$ and the corresponding wave number $k_{y,\max}$ lie between the small and large $\Delta '$ regimes. Approximate relations for $\gamma_{L,\max}$ and $k_{y,\max}$ can thus be found by balancing Eqs. (\ref{DR2_high_Deltap}) and (\ref{DR2_small_Deltap}). This gives
\begin{equation}
{\Delta '}_{\max} \sim \left( 2 \pi^2 \right)^{1/3} {d_e}^{-2/3} {\rho_\tau}^{-1/3} ,
\end{equation}
which leads to $k_{y,\max}$ if a known relationship between ${\Delta '}_{\max}$ and $k_{y,\max}$ exists, as is the case of the equilibrium described in Sec. \ref{sec3}. We obtain a great analytic simplification by adopting the approximation $\Delta ' \approx {{15}/{k_y^2}}$, which gives a good representation of Eq. (\ref{Deltap_ky}) for $k_y \lesssim 1$, as we expect to be the case for $k_y = k_{y,\max}$. Therefore, for the fastest growing mode we obtain the relations
\begin{equation}
k_{y,\max} \sim {\sqrt {15}} \left( 2 \pi^2 \right)^{-1/6} {d_e}^{1/3} {\rho_\tau}^{1/6} , \label{k_y_max}
\end{equation}
\begin{equation}
\gamma_{L,\max} \sim {\sqrt {15}} {\left( {\frac{2}{\pi^4 }} \right)^{1/6}} {{B_{y,eq}^\prime}} {d_e}^{2/3} {\rho_\tau}^{5/6} . \label{gamma_L_max}
\end{equation}
From the results of the gyrofluid simulations shown in Fig. \ref{fig2} we find that relation (\ref{k_y_max}) provide a very good estimate of $k_{y,\max}$, with a discrepancy never larger than $4\%$, while relation (\ref{gamma_L_max}) slightly overestimate the numerical results by a factor between 1.6 and 2.

We note that the effect of out-of-plane ion compressibility is retained in the gyrofluid model, thereby enabling the description of ion acoustic waves, which are not treated in the analytic theory discussed so far. In order to assess their role in the reconnection dynamics, we also consider the case in which ion acoustic waves are removed from the gyrofluid model by taking the limit $d_i \to \infty$. Eq. (\ref{e2}) then becomes
\begin{equation}
\frac{\partial u_i}{\partial t}+[\Phi,u_i]=0 , \label{eq0_3field}
\end{equation}
which imply that the out-of-plane velocity of the ion guiding centers remains unchanged as time
advances if $u_i = 0$ at $t=0$, as is the case of the equilibrium configuration specified in Sec. \ref{sec3}. As a consequence, the $z$-component of Amp\`ere's law reduces to $\nabla_\bot^2 \psi  =  - j = {u_e}$. Hence Eqs. (\ref{e1}), (\ref{e3}) and (\ref{e4}) become
\begin{equation}
\frac{\partial n_i}{\partial t}+[\Phi,n_i]=0 , \label{eq1_3field}
\end{equation}
\begin{equation}
\frac{\partial n_e}{\partial t}+[\phi,n_e] = [\psi, \nabla_\bot^2 \psi] ,\label{eq2_3field}
\end{equation}
\begin{equation}
\frac{\partial }{{\partial t}}\left( {\psi - d_e^2{\nabla_\bot^2 \psi}} \right)+[\phi,\psi-d_e^2{\nabla_\bot^2 \psi}] = - \rho_s^2 [\psi, n_e] ,
\label{eq3_3field}
\end{equation}
which are the same evolution equations of the three-field gyrofluid model \cite{WHM_2009} investigated in Refs. \cite{GTW_2010,TWG_2010}, where ion acoustic wave dynamics was neglected. Therefore, setting $d_i = 10^{6}$ we find a closer agreement with the growth rates obtained from Eq. (\ref{Porcelli_DR2}), as it is shown in Fig. \ref{fig4}. Moreover, a comparison between the gyrofluid growth rates in Figs. \ref{fig2} and \ref{fig4} allow us to quantify the effect of the ion acoustic waves on the tearing mode instability. We find that ion compressibility effects lead to a reduction of the growth rate over the whole range of linearly unstable wave numbers, with a greater impact in the intermediate $\Delta '$ regime for the largest $\beta$ value considered here. Even so, the discrepancy between the cases with and without ion compressibility effects is never larger than $8\%$. Ion acoustic wave coupling should become important for $\beta \sim 1 $, which however do not belong to the regime of validity of both the gyrofluid model and the analytic theory.

\section{Nonlinear Phase} \label{sec5}

In the investigation of the nonlinear evolution of the reconnecting instability, we restrict ourselves to the strongly unstable regime (large $\Delta '$), which is relevant to the general problem of fast magnetic reconnection. According to this choice, we set $L_y=4\pi$, which leads to $\Delta ' = 59.9$ for the longest wavelength mode in the system $k_y = {{2\pi}/{L_y}} = {{1}/{2}}$. We again consider high-temperature plasmas characterized by $\beta \gg 2{{m_e}/{m_i}}$, and we study the effects of the plasma parameters ($d_{e,i} $, $\rho_{s,i}$) on the reconnection dynamics. Since the model of Eqs. (\ref{e1})-(\ref{e6}) is dissipationless, we stop our simulations at a time when the microscopic structures associated with the reconnection process have become so narrow that they can no longer be resolved by our truncated Fourier expansion.

\begin{figure}[h!]
\begin{center}
\includegraphics[bb = 14 16 417 664, width=8.6cm]{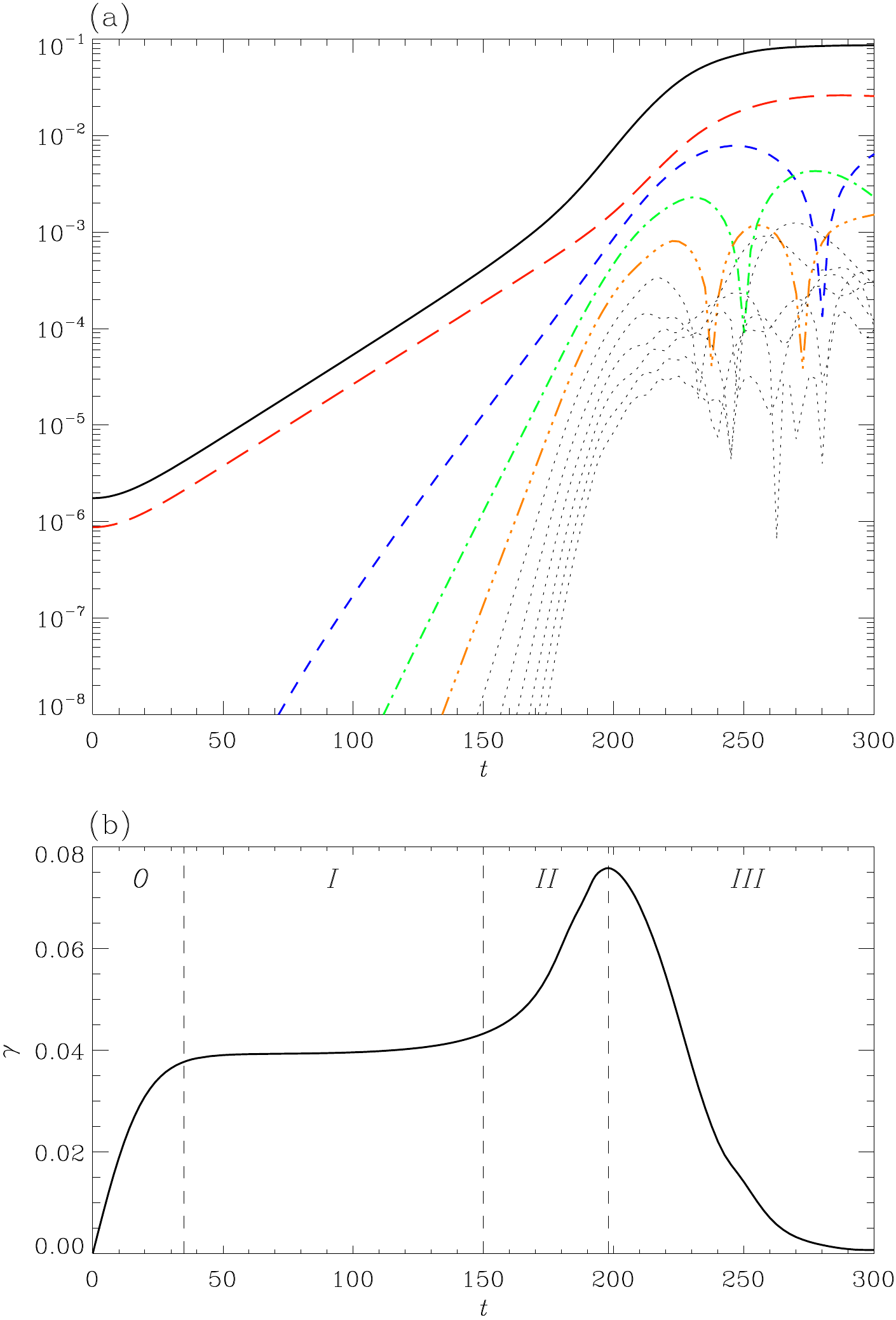}
\end{center}
\caption{(Color online) (a) Semi-log plot of the time evolution of the reconnected flux at the $X$-point $\delta {\psi_X}$ (black solid line) and the first 10 modes: $m=1$ (red long-dashed line), $m=2$ (blue short-dashed line), $m=3$ (green dashed-dotted line), $m=4$ (orange three dot-dashed line), $m=5,...,10$ (black dotted lines). Plasma parameters of this simulation are: $d_e = 0.2$, $d_i = 2$, $\rho_s = 0.4$, $\rho_i = 0.8$. The system size in the $y$-direction is $L_y=4\pi$, therefore $\Delta' \approx 59.9$ for the longest wavelength mode in the system. (b) Effective growth rate of the reconnecting instability, $\gamma = d(\ln \delta \psi_X)/dt$, as a function of time. After an initial transient ($0$), three main stages can be identified: the linear phase during which the growth rate is exponential ($I$), the super-exponential phase ($II$), and finally the saturation phase during which the growth rate slow down to zero as the reconnection is completed ($III$).}
\label{fig5}
\end{figure}

Fig. \ref{fig5} shows the evolution of magnetic reconnection in the strongly unstable collisionless regime for the following plasma parameters: $d_e = 0.2$, $d_i = 2$, $\rho_s = 0.4$, $\rho_i = 0.8$. In panel (a) it is shown the time evolution of the reconnected flux at the $X$-point, $\delta {\psi_X} = \left| {\psi (0,0,t) - {\psi _{eq}}(0)} \right|$, and that of the first ten modes, $\int {dx\delta {\psi _m}(x,t)}$. The $m>1$ modes develop due to the coupling of the mode initially exited ($m=1$). Indeed, at $t = 150$ we find that the growth rate for the $m > 1$ modes shown here is ${\gamma_{m > 1}} \approx m{\gamma_{m = 1}}$, in agreement with the predictions based on the quasilinear theory. Panel (b) shows the effective growth rate of the reconnecting instability, $\gamma = d(\ln \delta \psi_X)/dt$, as a function of time. From this plot we can clearly see that after an initial transient ($0$), magnetic reconnection evolves through three different stages: the linear phase ($I$), scaling as ${e^{{\gamma}t}}$, with $\gamma \approx 0.0401$, followed by the faster-than-exponential phase ($II$), during which the effective growth rate increases up to a peak value $\gamma \approx 0.0751$, and finally the saturation period ($III$) in which the growth rate slows down to zero as the reconnection is completed. We observe that the saturation occurs in spite of the energy conservation property of the Hamiltonian system. This happens because while the reconnecting instability develops, part of the magnetic energy is transferred from the large spatial scales towards the small scales, which are averaged out when the new coarse-grained stationary magnetic configuration is established \cite{GCPP_2001}.

\begin{figure}[h!]
\begin{center}
\includegraphics[bb = 14 10 436 508, width=8.6cm]{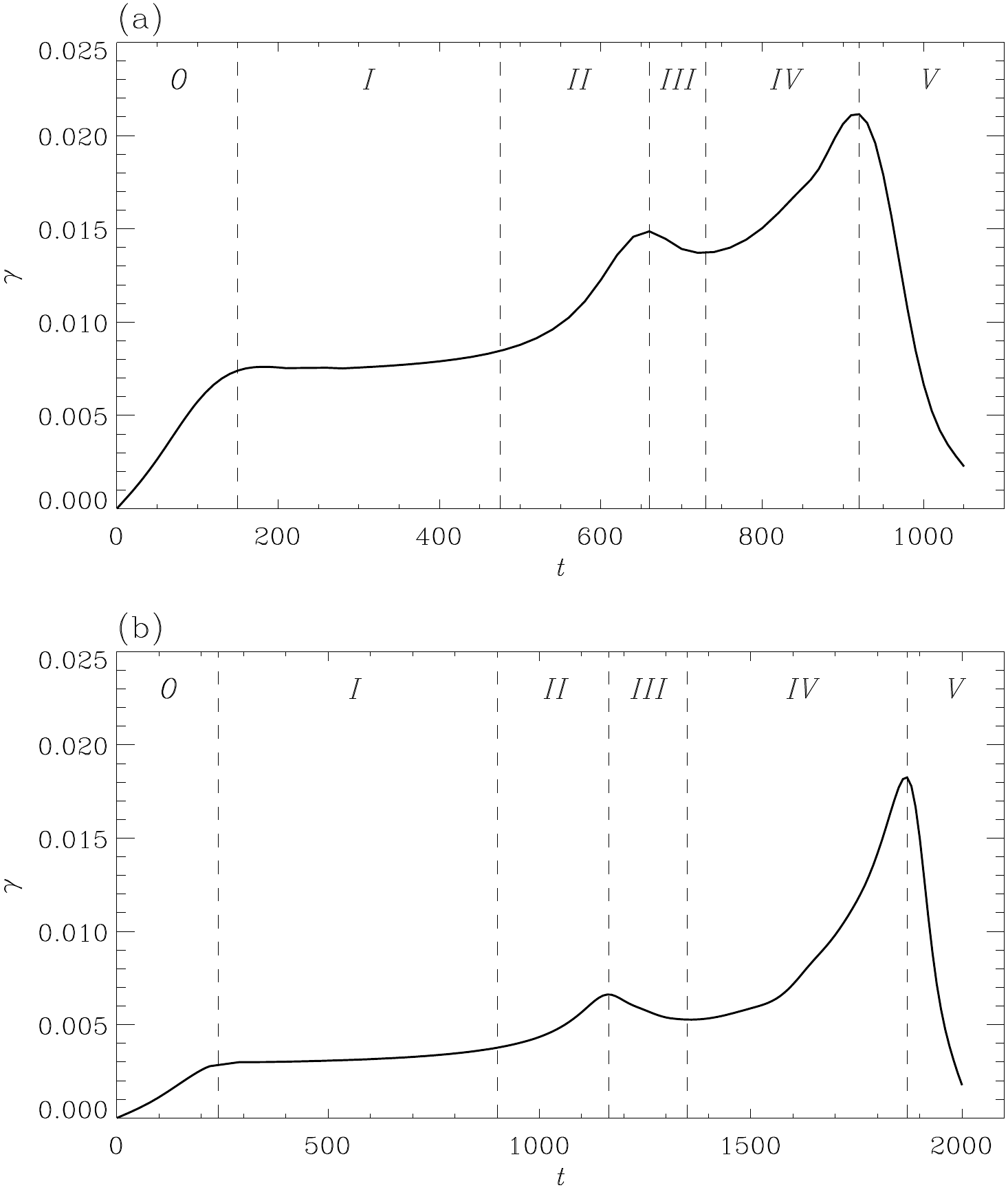}
\end{center}
\caption{Effective growth rate of the reconnecting instability, $\gamma = d(\ln \delta \psi_X)/dt$, as a function of time, for (a) $d_e = 5 \times {10^{-2}}$, $d_i = 0.5$, $\rho_s = 0.1$, $\rho_i = 0.2$, and (b) $d_e = 2.5 \times {10^{-2}}$, $d_i = 0.25$, $\rho_s = 5 \times {10^{-2}}$, $\rho_i = 0.1$. Both cases have the same ${{m_i}/{m_e}}$, $\beta_e$, $\beta_i$, $T_e$, $T_i$, and equilibrium configuration as in Fig. \ref{fig5}. After the initial transient ($0$), the reconnecting instability is seen here to evolve through five main stages: the linear phase ($I$), the first faster-than-exponential phase ($II$), the stall phase during which the growth rate slow down ($III$), the second faster-than-exponential phase characterized by a strong enhancement of the growth rate ($IV$), and the saturation phase ($V$).}
\label{fig6}
\end{figure}

A similar evolution of the growth rate was presented for the first time in a landmark paper by Aydemir \cite{A_1992}. In his work he considered the effects of finite $\rho_s$, but not those related to $\rho_i$ since he focused on the limit ${{T_i}/{T_e}} \to 0$. 
Subsequent studies \cite{OP_1993,WB_1993,RZ_1996,CGPPS_1998,Porc_2002,BGN_2005,Hirota_13} have confirmed that in the strongly unstable regime, the reconnecting instability undergoes one nonlinear acceleration with an instantaneous growth rate that is faster-than-exponential in time also when the nonlocal effects related to $\rho_i$ are taken into account \cite{GCPP_2000,DMPC_2011,BS_12,CGTW_2012}. In the following, extending our preliminary results \cite{CWG_2012}, we will show how this picture changes when considering hot ions ($\rho_i \gtrsim d_e$) and also a diffusion region thickness (both the electron and ion diffusion regions) that is effectively much smaller than the equilibrium magnetic field scale length (${d_e},\,{\rho_\tau} \ll L$), as is expected to be the case in most of space and laboratory plasmas \cite{PBR_1996,CS_2012}. Indeed, decreasing $d_e$ while keeping constant the values of ${{m_i}/{m_e}}$, $\beta_e$, $\beta_i$, $T_e$, $T_i$, we find that the nonlinear evolution of collisionless magnetic reconnection shows a novel behaviour, as shown in Fig. \ref{fig6}. Nonlinearly, the instantaneous growth rate is characterized by two distinct phases of strong increase, separated by a stall phase in which the growth rate decreases. Furthermore, the enhancement of the peak effective growth rate over its linear value increases with decreasing $d_e$ values, as can be seen by comparing Figs. \ref{fig6}(a) and \ref{fig6}(b). We will come back later on this point, and we focus now on the nonlinear evolution of the reconnection process.

\begin{figure}[h!]
\begin{center}
\includegraphics[bb = 14 10 436 508, width=8.6cm]{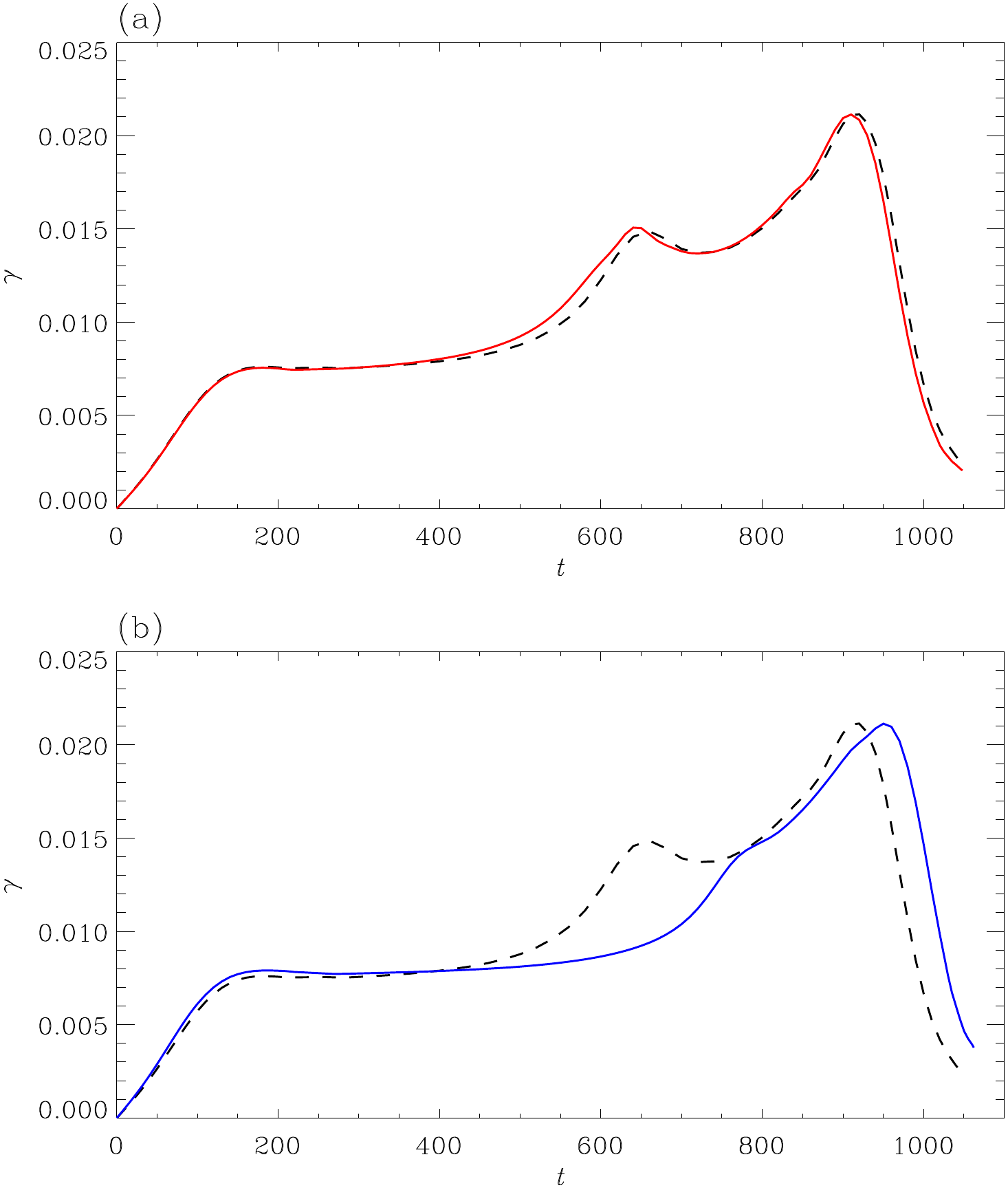}
\end{center}
\caption{(Color online) Effective growth rate of the reconnecting instability, $\gamma = d(\ln \delta \psi_X)/dt$, as a function of time, for (a) $\rho_s = 10^{-5}$, $\rho_i = 0.2236$ (red line), and (b) $\rho_s = 0.2236$, $\rho_i = 10^{-5}$ (blue line). The equilibrium configuration and the other plasma parameters are the same as in Fig. \ref{fig6}(a). The effective growth rate for the case with $\rho_s = 0.1$, $\rho_i = 0.2$ (black dashed line) is shown here for comparison. All these cases are characterized by the same value of ${\rho_\tau } = {\left( {\rho_s^2 + \rho_i^2} \right)^{1/2}}$.}
\label{fig7}
\end{figure}

\begin{figure}[h!]
\begin{center}
\includegraphics[bb = 13 7 336 679, width=8.6cm]{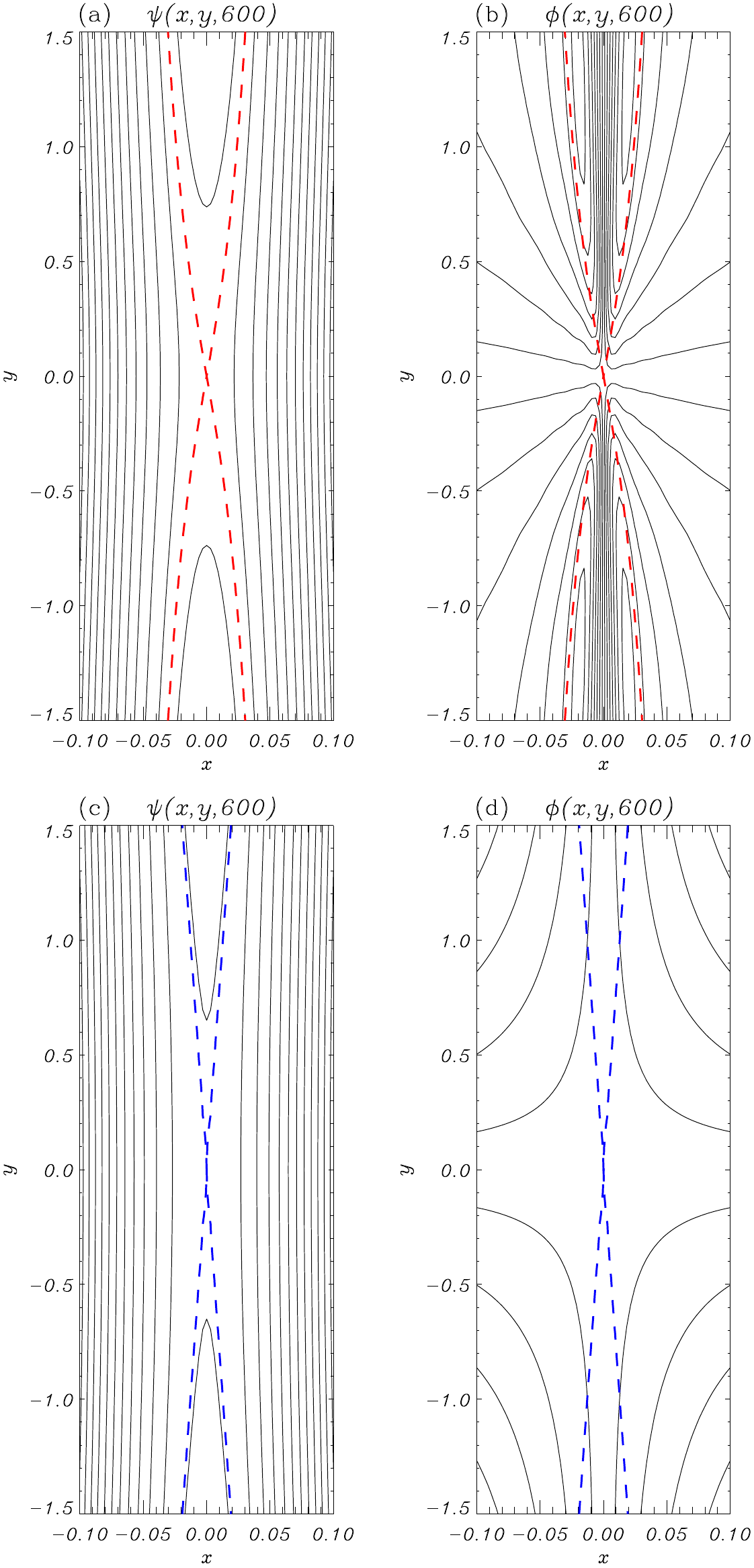}
\end{center}
\caption{(Color online) From the simulations shown in Fig. \ref{fig7}, isolines of the in-plane magnetic flux function $\psi$ and the electrostatic potential $\phi$ at $t=600$ for (top) $\rho_s = 10^{-5}$, $\rho_i = 0.2236$ and (bottom) $\rho_s = 0.2236$, $\rho_i = 10^{-5}$. For clarity, only a small portion of the computational domain is plotted with an altered aspect ratio. The magnetic island separatrix at the corresponding time have been superimposed (dashed lines). The full width of the magnetic island is $w \approx 2 d_e = 0.1$ in both cases, but for hot ions the separatrix is characterized by a greater opening close to the $X$-point.}
\label{fig8}
\end{figure}

\begin{figure}[h!]
\begin{center}
\includegraphics[bb = 14 7 436 565, width=8.6cm]{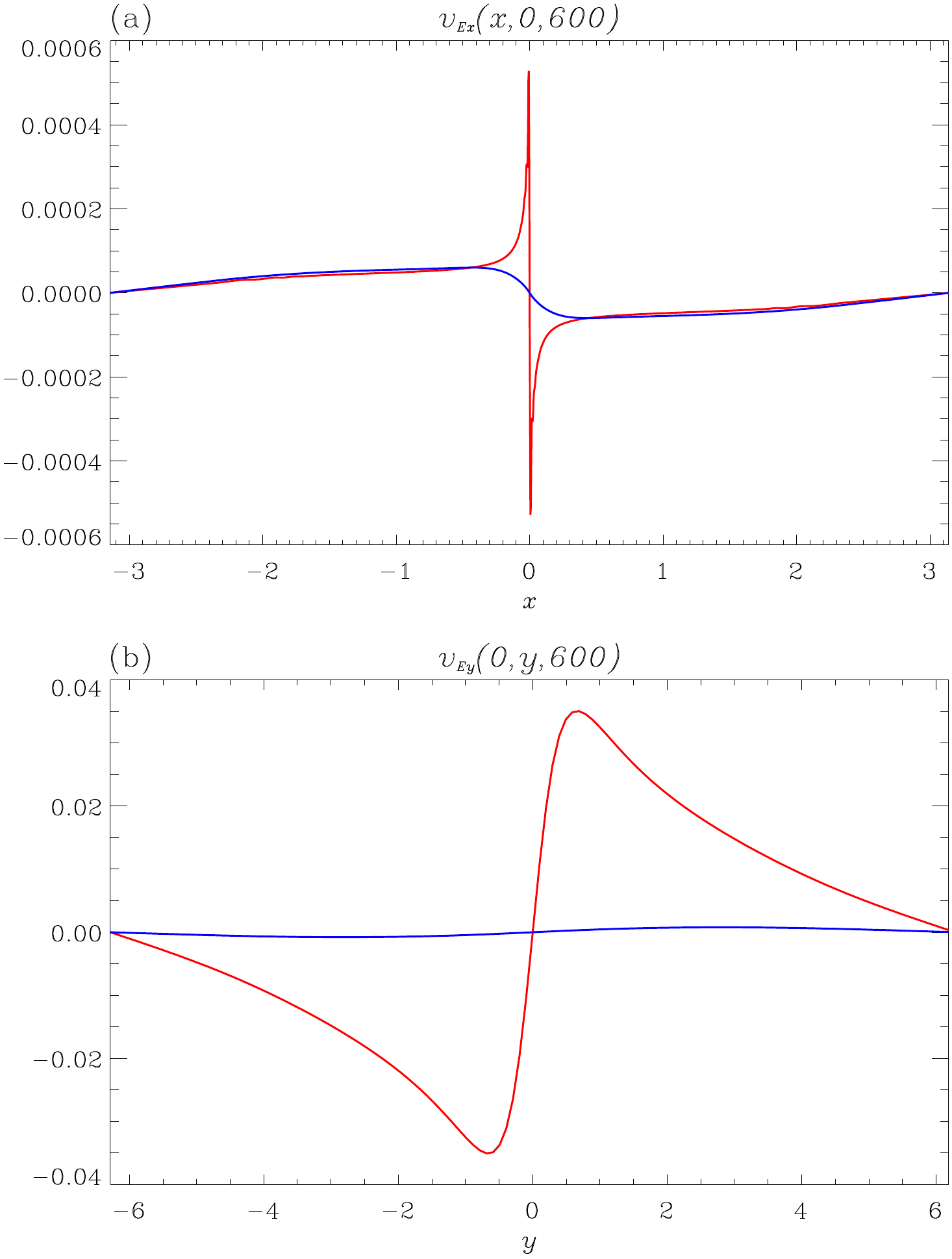}
\end{center}
\caption{(Color online) From the same simulations as in Fig. \ref{fig7}, profiles of the (a) $x$-component of the ${\bf{E}} \times {\bf{B}}$ flow velocity at $y=0$, $t=600$ and the (b) $y$-component of the ${\bf{E}} \times {\bf{B}}$ flow velocity at $x=0$, $t=600$. The red line refers to $\rho_s = 10^{-5}$, $\rho_i = 0.2236$, whereas the blue line refers to $\rho_s = 0.2236$, $\rho_i = 10^{-5}$. Note that, in  the hot ion case, $\max \left| {{v_{Ex}}\left( {x,0,600} \right)} \right|$ is about one order of magnitude higher than in the cold ion case. An even larger difference between the hot and cold ion cases is found for $\max \left| {{v_{Ey}}\left( {0,y,600} \right)} \right|$.}
\label{fig9}
\end{figure}

To distinguish ion gyration effects from those related to the electron out-of-plane compressibility, $\rho_i$ and $\rho_s$ are varied while keeping ${\rho_\tau } = {\left( {\rho_s^2 + \rho_i^2} \right)^{1/2}} = \rm{const} $. With the choices $\left( {{\rho_i},{\rho_s}} \right) = \left\{ {\left( {0.2236,10^{-5}} \right),\left( {10^{-5},0.2236} \right)} \right\}$ we obtain the same $\rho_\tau$ as in Fig. \ref{fig6}(a). The corresponding evolutions of the instantaneous growth rate are shown in Figs. \ref{fig7}(a) and \ref{fig7}(b). The first acceleration phase is present only in the hot ion case, and is found to begin at $t \approx 450$, which corresponds to a full island width $w \approx d_e = 0.05$. Conversely, when ions are cold the early acceleration is absent. This different behaviour can be explained by looking at the field structures around the $X$-point (a saddle point in $\psi$) for the two cases. Fig. \ref{fig8} shows a zoom around the $X$-point of the isolines of the fields $\psi$ and $\phi$ at $t = 600$. At this stage of the reconnection process the island widths are of the same order, but the hot ion case is characterized by a greater opening of the magnetic island separatrix that allows for a wider outflow region \cite{CGTW_2012}. Moreover, in the large ion Larmor radius case ${\bf{v}}_E = {\bf{\hat z}} \times \nabla \phi$ converges toward the $X$-point leading to much smaller structures. Similar patterns of the field $\phi$ were shown also in Ref. \cite{GCPP_2000}, however, the length scale separation was not sufficient to identify more than one nonlinear acceleration. To make the comparison between the hot and cold ion cases more quantitative, in Fig. \ref{fig9} it is shown the magnitude of the $\bf{E} \times \bf{B}$ flow velocities along the inflow and outflow directions across the $X$-point. For the hot ion case it is found that $\max \left| {{v_{Ex}}\left( {x,0,600} \right)} \right|$ is about one order of magnitude higher than in the cold ion case, and an even larger difference is found for $\max \left| {{v_{Ey}}\left( {0,y,600} \right)} \right|$. The change in the behaviour of the $\bf{E} \times \bf{B}$ flow velocities leads to the different instantaneous growth rate for hot and cold ions at this stage of the reconnection process. Therefore, the first acceleration phase that appears when ions are hot can be explained by looking at the spatial structures in the field $\phi$ in Fig. \ref{fig8}. 

\begin{figure}[h!]
\begin{center}
\includegraphics[bb = 13 7 346 495, width=8.6cm]{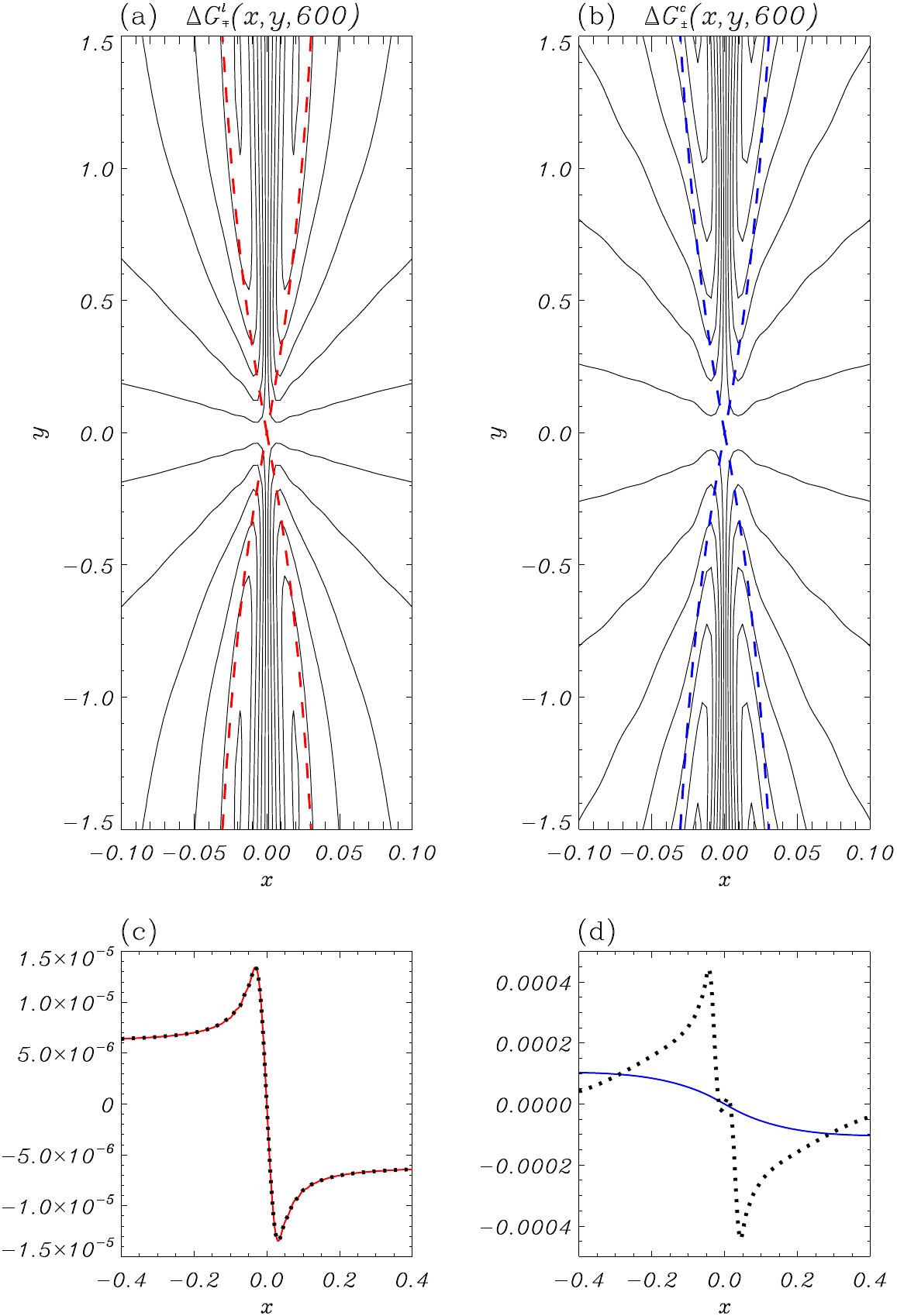}
\end{center}
\caption{(Color online) From the same simulations as in Fig. \ref{fig7}, isolines of (a) $\Delta G_\mp ^l = G_-^l - G_+^l$ at $t=600$ for $\rho_s = 10^{-5}$, $\rho_i = 0.2236$, and (b) $\Delta G_\pm ^c = G_+^c - G_-^c$ at $t=600$ for $\rho_s = 0.2236$, $\rho_i = 10^{-5}$. For clarity, only a small portion of the computational domain is plotted with an altered aspect ratio. The magnetic island separatrix at the corresponding time have been superimposed (dashed lines). Profiles of (c) $\phi$ (red solid line) and $\rho _i^2 \cdot {{\Delta G_ \mp ^l}/{(2{d_e}{\rho_s})}}$ (black dotted line) at $y=0.2$, $t=600$ for the hot ion case, and (d) $10 \cdot \phi$ (blue solid line) and ${{\Delta G_ \pm ^c}/{(2{d_e}{\rho_s})}}$ (black dotted line) at $y=0.2$, $t=600$ for the cold ion case. For clarity, only the interval $-0.4 \le x \le 0.4$ is plotted. Panel (c) confirms numerically the validity of relation (\ref{phi_hot}) in the large ion Larmor radius limit.}
\label{fig10}
\end{figure}

To understand the behaviour of the electrostatic potential it is useful to carry out some analytical considerations. We note that since the first acceleration phase occurs when the island width exceeds the thickness of the electron layer but not that of the ions, i.e. ${d_e} < {{w}/{2}} < {\rho_\tau}$, we can consider only the region $\left| x \right| < {\rho_\tau }$ around the $X$-point, where the motion of the ions is essentially their gyro-motion. Hence, the out-of-plane dynamics is determined only by the electrons, whose equations of continuity and motion in the $z$-direction are, respectively, Eqs. (\ref{eq2_3field}) and (\ref{eq3_3field}), closed by the quasi-neutrality condition
\begin{equation}
{n_e} = \frac{{\left( {{\Gamma_0} - 1} \right)}}{{\rho_i^2}}\phi . \label{e6_new}
\end{equation}
In the limit $k_\bot ^2\rho _i^2 \ll 1$, Eq. (\ref{e6_new}) expresses the fact that the density is equal to the $\bf{E} \times \bf{B}$ vorticity ${n_e} = \nabla_ \bot^2 \phi$, in which case Eqs. (\ref{eq2_3field}) and (\ref{eq3_3field}) reduce simply to
\begin{equation}
\frac{{\partial \nabla _ \bot ^2\phi }}{{\partial t}} + \left[ {\phi ,\nabla _ \bot ^2\phi } \right] = \left[ {\psi ,\nabla _ \bot ^2\psi } \right] ,
\end{equation}
\begin{equation}
\frac{\partial }{{\partial t}}\left( {\psi  - d_e^2\nabla _ \bot ^2\psi } \right) + \left[ {\phi ,\psi  - d_e^2\nabla _ \bot ^2\psi } \right] = - \rho _s^2\left[ {\psi , \nabla _ \bot ^2\phi} \right] ,
\end{equation}
that can be cast in the following Lagrangian conservative form \cite{CGPPS_1998,GCPP_2001}
\begin{equation}
\frac{{\partial G_ \pm ^c}}{{\partial t}} + \left[ {\phi _ \pm ^c,G_ \pm ^c} \right] = 0 , \label{G_cold}
\end{equation}
where 
\begin{eqnarray}  
G_ \pm ^c = \psi  - d_e^2\nabla _ \bot ^2\psi  \pm {d_e}{\rho _s}\nabla _ \bot ^2\phi , \qquad \phi _ \pm ^c = \phi  \pm \frac{{{\rho _s}}}{{{d_e}}}\psi . 
\end{eqnarray} 
On the other hand, in the limit $k_\bot ^2\rho _i^2 \gg 1$, at the leading order Eq. (\ref{e6_new}) reduces to the relation $n_e = - {{\phi}/{\rho_i^2}}$, and the system (\ref{eq2_3field}) - (\ref{eq3_3field}) becomes
\begin{equation}
\frac{{\partial \phi }}{{\partial t}} = \rho _i^2\left[ {\nabla _ \bot ^2\psi ,\psi } \right] ,
\end{equation}
\begin{equation}
\frac{\partial }{{\partial t}}\left( {\psi  - d_e^2\nabla _ \bot ^2\psi } \right) + \left[ {\phi ,\psi  - d_e^2\nabla _ \bot ^2\psi } \right] = \frac{{\rho _s^2}}{{\rho _i^2}}\left[ {\psi ,\phi } \right] ,
\end{equation}
which can as well be cast in Lagrangian conservative form
\begin{equation}
\frac{{\partial G_ \pm ^l}}{{\partial t}} + \left[ {\phi _ \pm ^l,G_ \pm ^l} \right] = 0 , \label{G_hot}
\end{equation}
where 
\begin{eqnarray}  
G_ \pm ^l = \psi  - d_e^2\nabla _ \bot ^2\psi  \mp \frac{{{d_e}{\rho _s}}}{{\rho _i^2}}\phi , \qquad \phi _ \pm ^l = \phi _ \pm ^c . 
\end{eqnarray} 
Therefore, the structure of the electrostatic potential around the $X$-point can be linked to $G_ \pm ^c$ in the cold ion limit, and to $G_ \pm ^l$ in the large ion Larmor radius limit. Indeed, subtracting the invariants $G_ \pm ^c$ we obtain
\begin{eqnarray}  
 \phi = \nabla_ \bot ^{-2}\left( {\frac{{G_+^c - G_-^c}}{{2{d_e}{\rho_s}}}} \right) , \label{phi_cold}
\end{eqnarray} 
whereas, from the difference beetween the invariants $G_ \pm ^l$ we obtain
\begin{eqnarray}  
\phi = \rho_i^2\frac{{G_-^l - G_+^l}}{{2{d_e}{\rho_s}}} . \label{phi_hot}
\end{eqnarray} 
Hence, from relation (\ref{phi_cold}) we can infer that in the cold ion limit $\phi$ is smoothed with respect to $\Delta G_\pm ^c = G_+^c - G_-^c$, while relation (\ref{phi_hot}) shows that in the large ion Larmor radius limit $\phi$ is proportional to $\Delta G_\mp ^l = G_-^l - G_+^l$. Since the topological constraints set by Eqs. (\ref{G_cold}) and (\ref{G_hot}) force the $G$-family Lagrangian invariants to develop small scale structures in both the hot (${\rho_s} \ne 0$) and cold (${\rho_s} \to 0$) electron regimes, as shown in Ref. \cite{CGPPS_1998} and in several subsequent works, it is now evident that when ions are hot the field $\phi$ retains the small scale structures of $G_-^l - G_+^l$. This is clearly shown in Figs. \ref{fig10}(a) and \ref{fig10}(b), where both $\Delta G_\mp ^l$ and $\Delta G_\pm ^c$ exhibit small scale structures, which are reflected in the field $\phi$ only in the hot ion case, as can be seen by comparing Figs. \ref{fig10}(a) and \ref{fig10}(b) with Figs. \ref{fig8}(b) and \ref{fig8}(d). As a further confirmation of the previous analytical considerations, in Fig. \ref{fig10}(c) are shown the profiles of $\phi$ (red solid line) and $\rho _i^2 \cdot {{\Delta G_ \mp ^l}/{(2{d_e}{\rho_s})}}$ (black dotted line) at $y=0.2$, $t=600$, for the hot ion case. The two lines are essentially indistinguishable from one another, thus confirming the validity of relation (\ref{phi_hot}) in the limit $k_\bot ^2\rho _i^2 \gg 1$. In Fig. \ref{fig10}(d), a similar comparison for the cold ion case shows that $\phi$ (we have plotted $\phi \cdot 10$ for clarity) is smoothed with respect to ${{\Delta G_ \pm ^c}/{(2{d_e}{\rho_s})}}$, as predicted by relation (\ref{phi_cold}).

\begin{figure}[h!]
\begin{center}
\includegraphics[bb = 9 1 285 400, width=8.6cm]{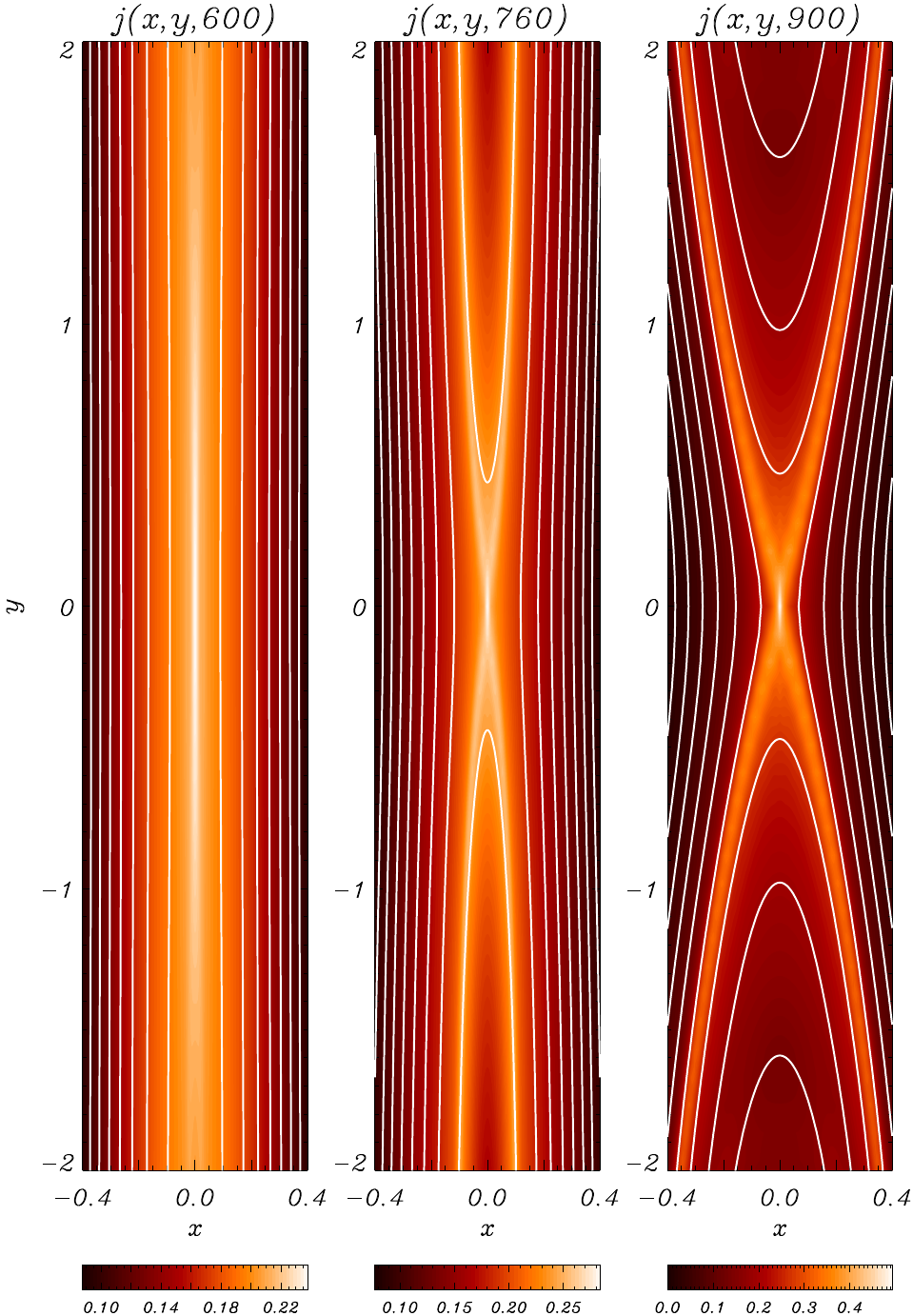}
\end{center}
\caption{(Color online) From the simulation shown in Fig. \ref{fig6}(a), blowup around the $X$-point of the out-of-plane current density with magnetic field lines (white lines) superimposed at (left frame) $t=600$, during the first acceleration phase, at (central frame) $t=760$, at the beginning of the second acceleration phase, and at (right frame) $t=900$, well into the second acceleration phase. the full width of the magnetic island is $w=0.097$ at $t=600$, $w=0.266$ at $t=760$, and $w=0.920$ at $t=900$.}
\label{fig11}
\end{figure}

After the first acceleration phase induced by the ion Larmor gyration, the instantaneous growth rate decreases only to undergo a strong enhancement when a Petschek-like configuration arises due to finite ${\rho_\tau}$ values. This is shown in Fig. \ref{fig11}, where the out-of-plane current density and the magnetic field lines are plotted at three different times of the simulation shown in Fig. \ref{fig6}(a). The early nonlinear phase is characterized by a thin current sheet, while the second acceleration phase occurs when the out-of-plane current density from the $X$-point opens, giving rise to a macroscopic outflow region that speeds up the reconnection process. A microscopic current sheet persists at the $X$-point with a width that shrinks in time as the reconnection proceeds. Indeed, the adopted gyrofluid model does not contain cutoff dissipative scale lengths. It is clear that in a real plasma this tendency toward a singular behavior would be limited by additional physics not taken into account in the model, such as, for instance, electron Larmor radius effects or velocity space instabilities. However, performing a simulation with double the resolution in the $x$-direction showed that the reconnection rate is not affected by the size of the numerical dissipation at scales well below the electron skin depth, whereas the quantity that is most sensitive to numerical dissipation, the peak electron velocity at the $X$-point, increases by less than $6\%$ when the resolution is doubled. Previous works \cite{OP_1995,GPPC_1999} explained this finding as a consequence of the fact that the nonlinear microscales narrower than the electron skin depth carry a negligible current with respect to that distributed over a width of order $d_e$. Similar conclusions were also obtained in the context of electron MHD (see, e.g., Ref. \cite{CSZ_2007}), where it was found that the reconnection rate becomes independent of the dissipation coefficient in the limit of a very small magnetic dissipation.

We observe that an X-type magnetic field configuration (not shown here) develops also during the nonlinear acceleration of the cold ion case shown in Fig. \ref{fig7}(b). This is due to electron temperature effects, as pointed out in previous works \cite{A_1992,KDW_1995,RZ_1996,MB_1996,CGPPS_1998,GPPC_1999,RDDS_2001}.

\begin{figure}[h!]
\begin{center}
\includegraphics[bb = 14 10 436 237, width=8.6cm]{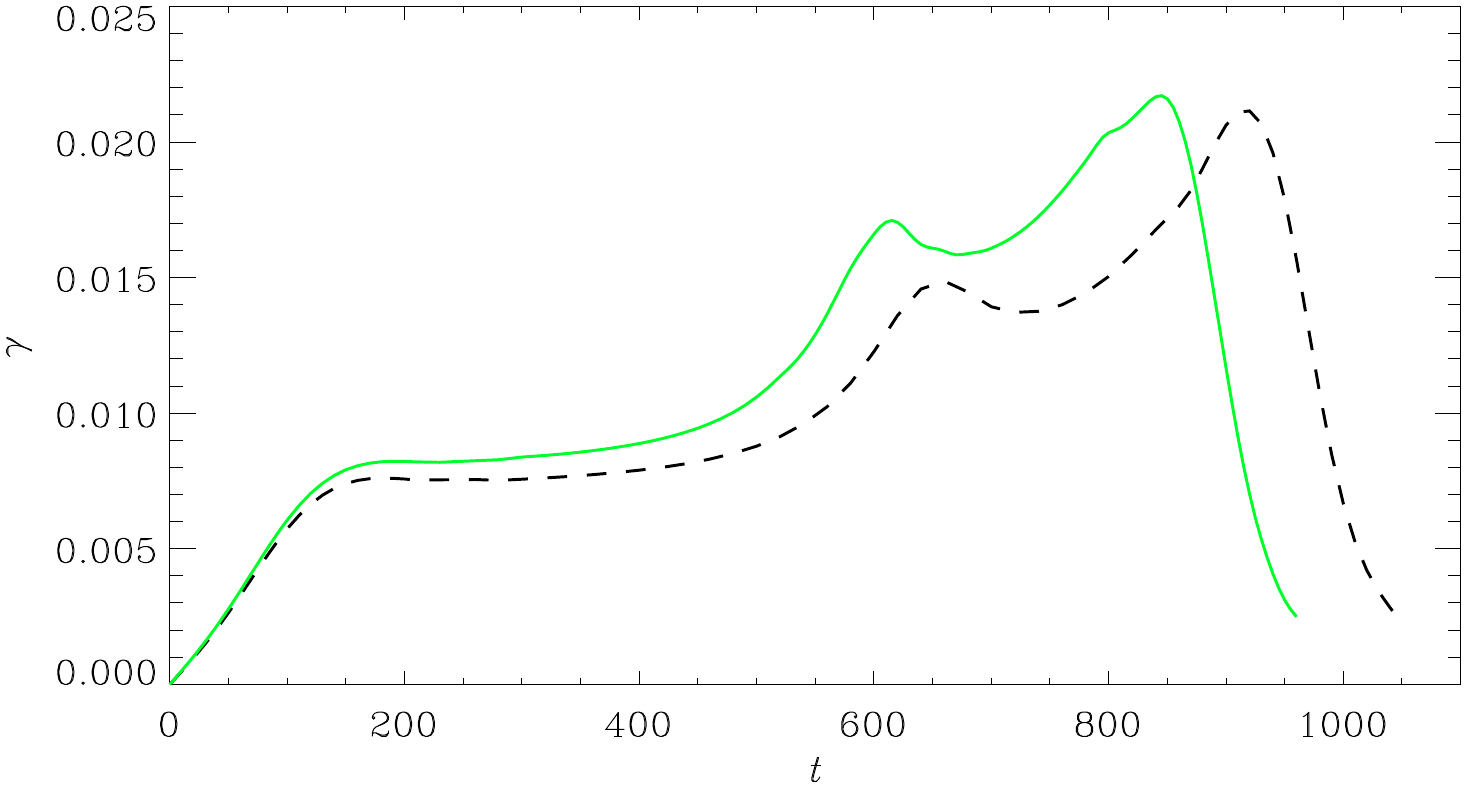}
\end{center}
\caption{(Color online) Effective growth rate of the reconnecting instability, $\gamma = d(\ln \delta \psi_X)/dt$, as a function of time, for $d_i = 10^{6}$ (this choice identifies the case $d_i \to \infty$), which has the effect of eliminating ion acoustic waves (green solid line). The equilibrium configuration, as well as the other plasma parameters, are the same as in Fig. \ref{fig6}(a), whose effective growth rate is shown here for comparison (black dashed line).}
\label{fig12}
\end{figure}

Note that for the cases investigated here, characterized by $\beta = 2 {{c_s^2}/{v_A^2}} < 2 \left({{m_e}/{m_i}}\right)^{1/4}$ and constant equilibrium density, the qualitative evolution of magnetic reconnection does not depend on ion acoustic wave dynamics. In fact, Fig. \ref{fig12} shows that also in the limit $d_i \to \infty$ the reconnection process undergoes two phases of strong increase of the instantaneous growth rate. Moreover, ion acoustic wave dynamics is found to be stabilizing also in the nonlinear phase since the effective growth rate of the reconnecting instability is higher in the regime without ion compressibility effects than in the case with out-of-plane ion compressibility.

\begin{figure}[h!]
\begin{center}
\includegraphics[bb = 60 7 390 475, width=8.6cm]{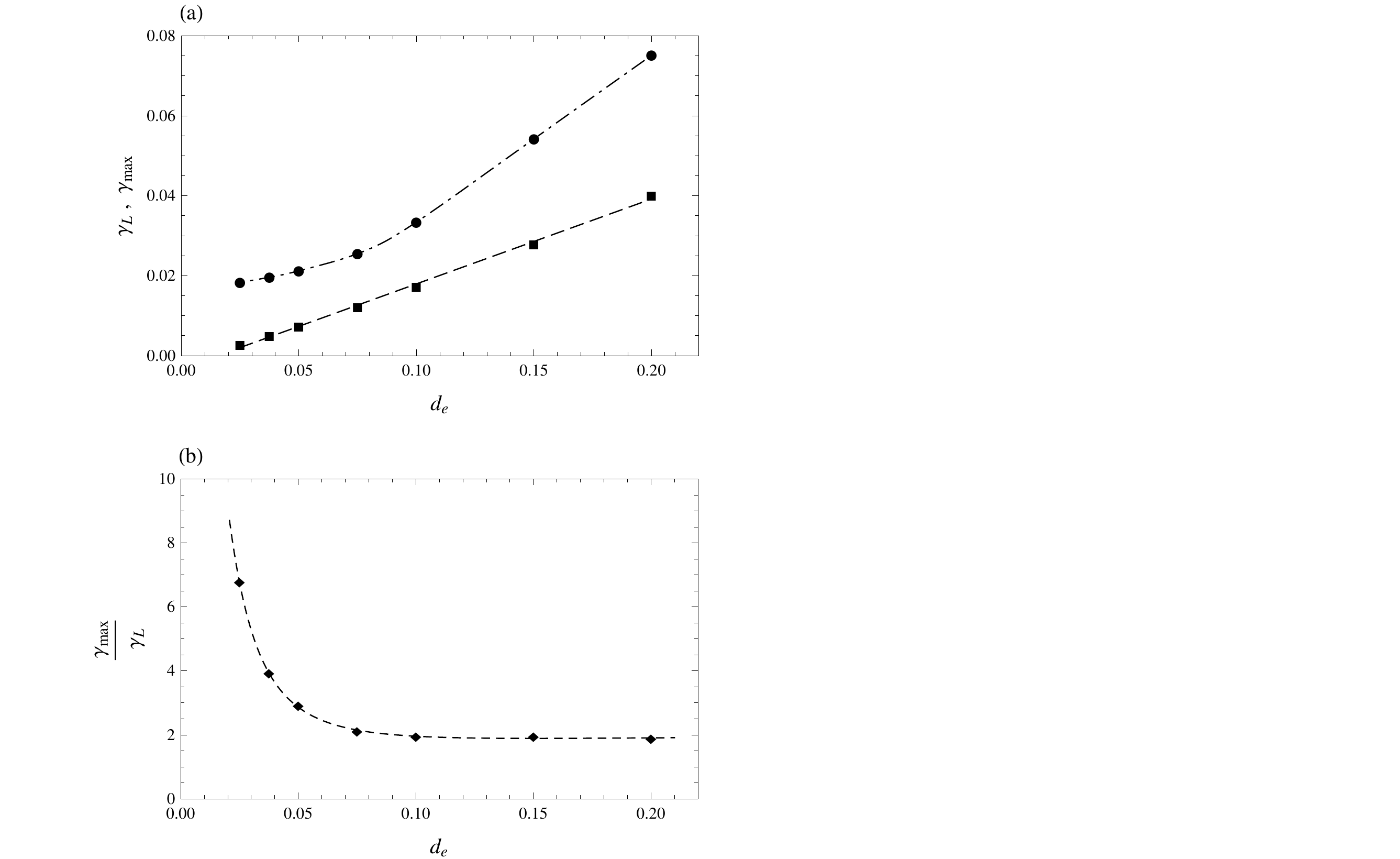}
\end{center}
\caption{(a) Scaling of the linear and peak effective growth rates with electron skin depth. The square data points are linear growth rates (long-dashed line), while circle data points are peak growth rates (dashed-dotted line). All runs are characterized by the same equilibrium configuration and the same values of ${{m_i}/{m_e}}$, $\beta_e$, $\beta_i$, $T_e$, $T_i$. (b) Ratio between the peak effective growth rate and the linear one as a function of $d_e$ (short-dashed line).}
\label{fig13}
\end{figure}

By comparing Figs. \ref{fig6}(a) and \ref{fig6}(b), we have previously observed that the enhancement of the maximum effective growth rate over its linear value increases with decreasing ratios of the electron skin depth to the equilibrium magnetic field scale length. A quantitative evaluation is shown in Fig. \ref{fig13}. Panel (a) shows the scaling of the linear and maximum effective growth rates with $d_e$, while in panel (b) it is shown their ratio as a function of $d_e$. We observe that all runs are characterized by the same equilibrium configuration and the same values of ${{m_i}/{m_e}}$, $\beta_e$, $\beta_i$, $T_e$, $T_i$. The growth rate of the linear phase, that corresponds approximately to the linear growth of the $m=1$ mode, scale linearly with the $d_e$ values examined here, which are such that $d_e \Delta ' \gtrsim 1$. This is in agreement with the linear theory. Indeed, from Eq. (\ref{DR2_high_Deltap}) we know that $\gamma_L \propto {d_e}^{1/3} {\rho_\tau}^{2/3}$, but since in the scaling of Fig. \ref{fig13} we have $\rho_s = c_1 d_e$ and $\rho_i = c_2 d_e$, with $c_1$ and $c_2$ constants, the proportionality relation translates to $\gamma_L \propto {\left( c_1^2 + c_2^2 \right)}^{1/3} {d_e} \propto d_e$. The peak effective growth rate ${\gamma_{\max}}$ decreases linearly with $d_e$ for $0.1 \lesssim d_e \ll L$, whereas it asymptotes to a constant value for lower $d_e$. Therefore, ${\gamma_{\max}}$ becomes {\it weakly} dependent on $d_e$ (i.e. the mechanism that breaks the frozen-in condition) when the thickness of both the electron and ion diffusion regions (which scale like $d_e$ and ${\rho_\tau}$, respectively) are effectively much smaller than the equilibrium magnetic field scale length (${d_e},\,{\rho_\tau} \ll L$). Note that in the limit of no-guide field, $d_i$ replaces ${\rho_\tau}$ as the typical length scale of the ion diffusion region thickness \cite{Birn_2001,Shay_2004}. 

As a consequence of the behaviour of linear and peak effective growth rates, ${{\gamma_{\max}}/{\gamma_L}} \sim 2$ for $0.1 \lesssim d_e \ll L$, whereas for lower $d_e$ values the peak effective growth rate exhibits a dramatic enhancement over its linear value, as shown in Fig. \ref{fig13}(b). However, it is important to point out that when the limit $d_e \Delta ' \ll  ({{{d_e}/{\rho_\tau}}})^{1/3}$ is reached the faster-than-exponential phase vanishes because in this case the nonlinear regime is characterized by very thin islands ($w \Delta ' \ll 1$) for which the constant $\psi$ approximation applies across the island. Therefore, a Rutherford-like phase follows the linear phase and the magnetic island saturates at a microscopic width \cite{DL_1977_PRL,Sydora_2001}.

In order to make contact with observations and previous theoretical works, we evaluate the peak reconnection rate by calculating the maximum out-of-plane electric field at the $X$-point. Indeed, in two-dimensional reconnection the breaking and rejoining of magnetic field lines can take place only at an $X$-point, and the reconnection rate ${E_{z,X}}$ is a measure of the temporal rate of change of magnetic flux that undergoes this process. Since the $z$-component of the electrostatic field vanishes at the resonant surface, from Faraday's law $ E_{z,X} = - d{\psi_X}/dt$, with $\psi_X = \psi (0,0,t)$. We recall that from relations (\ref{modsp}) the electric field is normalized to ${v_A}{B_0}$, but to facilitate comparison with previous works we renormalize the reconnection rate using $v_{A,up} B_{y0,up}$, where we choose $B_{y0,up} = \max \left| {{B_{y,eq}}} \right|$, that corresponds to the in-plane equilibrium magnetic field at $x \approx \pm 0.66$ from the rational surface, and $v_{A,up} = B_{y0,up}/{{\left( {{\mu_0}{n_0}{m_i}} \right)^{{1}/{2}}}}$. This choice is admittedly ad hoc, but nevertheless it is reasonable for the purpose of an estimate. The resulting peak reconnection rates from the simulations shown in Fig. \ref{fig13} are listed in table \ref{tab:table1}. For sufficiently large systems, which in this case means $L \gg {d_e},\,{\rho_s},\,{\rho_i}$ and $d_e \Delta ' \gtrsim 1$, Table I shows that (in dimensional units) $E_{z,X}^{\max} \sim 0.1 v_{A,up} B_{y0,up}$, in qualitative agreement with the results of the numerical simulations in Refs. \cite{RDDS_2001,Rogers_2007,CDS_2007,LSZ_2013} for fast magnetic reconnection with a large guide field. This peak reconnection rate is also consistent with observed fast energy release rates \cite{Edw_1986,Isobe_2005,Egedal_2007}. We stress again that when the diffusion region thickness is so thin that $d_e \Delta ' \ll  ({{{d_e}/{\rho_\tau}}})^{1/3}$, the reconnection becomes a slow diffusive process, consequently the reconnection rate drops until $E_{z,X}^{\max} \to 0$ as $d_e \to 0$.

\section{Summary} \label{sec6}

We have explored the linear and nonlinear evolution of magnetic reconnection phenomena in which the reconnecting component of the magnetic field is small compared to the total magnetic field strength. Adopting a gyrofluid model for collisionless plasmas, we have studied the effects of ion gyration, ion and electron compressibility, and electron inertia on the growth rate of the reconnecting instability. In the linear theory limit, we have compared the growth rates obtained from gyrofluid simulations with analytical calculations across the entire spectrum of linearly unstable wave numbers. Focusing on high-temperature plasmas characterized by $\beta \gg 2{{m_e}/{m_i}}$, we have found a good agreement between the theory and the simulations, even if for $d_e$ values not asymptotically small the gyrofluid growth rates have a weaker dependence on ${{T_i}/{T_e}}$ and $\beta_i$ than the analytic theory. Furthermore, we have shown that the inclusion of the ion acoustic wave dynamics have stabilizing effects in both cold and hot ion regimes.

In the investigation of the nonlinear evolution of the reconnecting instability we have focused on the strongly unstable regime (large $\Delta '$), which is relevant to the general problem of fast magnetic reconnection.
We have shown for the first time that the nonlinear evolution of the reconnection process undergoes a novel behaviour when ions are hot ($\rho_i \gtrsim d_e$) and the diffusion region thickness is effectively much smaller than the equilibrium magnetic field scale length (${d_e},\,{\rho_\tau} \ll L$), as is expected to be the case in most of space and laboratory plasmas \cite{PBR_1996,CS_2012}. Under these circumstances, magnetic reconnection undergoes two distinct acceleration phases characterized by a strong increase of the instantaneous growth rate. The first nonlinear acceleration is due to ion temperature effects. In fact, we have shown that the ion Larmor gyration is responsible for the development of strong gradients of the electrostatic potential close to the $X$-point, which in turn lead to large $\bf{E} \times \bf{B}$ flows that speed up the reconnection. After a stall phase in which the instantaneous growth rate decreases, a second acceleration phase begins due to both ion and electron temperature effects that allow the emergence of a Petschek-like configuration.
In the low-$\beta$ regimes considered here, i.e. when ${k_\parallel}{v_A} > {k_\parallel}{c_s}$, the out-of-plane ion compressibility does not change this qualitative picture. 

Finally, the peak effective growth rate of the reconnection process is found to increases dramatically over its linear value for sufficiently large systems. This is because the effective growth rate of the linear phase depends strongly on the microscopic plasma parameters, while the peak effective growth rate becomes {\it weakly} dependent on the electron inertia and the other microscopic parameters when $L \gg {d_e},\,{\rho_s},\,{\rho_i}$ and $d_e \Delta ' \gtrsim 1$. When these limits are fulfilled, the peak reconnection rate scale roughly as $E_{z,X}^{\max} \sim 0.1 v_{A,up} B_{y0,up}$, that is fast enough to explain observed fast energy release rates \cite{Edw_1986,Isobe_2005,Egedal_2007}.

\acknowledgments

It is a pleasure to acknowledge fruitful discussions with Ahmet Aydemir, Vladimir Lakhin, Homa Karimabadi, Giannis Keramidas, William Matthaeus, Anna Perona, Emanuele Tassi and Oliver Zacharias. One of us (L.C.) is grateful for the hospitality of the Institute for Fusion Studies at the University of Texas at Austin, where part of this work was done.
This work was supported by the European Community under the contracts of Association between EURATOM and ENEA and by the U.S. Department of Energy under Contract No. DE-FG02-04ER-54742. The views and opinions expressed herein do not necessarily reflect those of the European Commission.

\section*{APPENDIX: DISPERSION RELATION}

Let us consider a homogeneous equilibrium described by ${n_{{i,eq}}} = {n_{{e,eq}}} = \rm{const} $, $u_{i,eq} = 0$, ${\phi_{eq}} = 0$, and ${{\bf{B}}_{ \bot,eq}} = \nabla \psi_{eq} \times {\bf{\hat z}} = {B_{y0}}{\bf{\hat y}}$, with $B_{y0}$ a constant. If we assume that all the fields can be written as $\chi = \chi_{eq} + \delta \chi \left( {x,y,t} \right)$, where $\delta \chi$ represents small perturbations that behave like $\exp \left( {i{k_x}x + i{k_y}y - i\omega t} \right)$, the linearized versions of Eqs. (\ref{e1})-(\ref{e4}) for the Fourier components are:
\begin{equation}
i\omega {{\hat n}_i} - i{k_y}{{\hat u}_i} {B_{y0}} = 0 ,
\end{equation}
\begin{equation}
i\omega {\Gamma_0^{{1}/{2}}}\left( b \right) {\hat \psi}  + i\omega d_i^2{{\hat u}_i} - i{k_y} {\Gamma_0^{{1}/{2}}}\left( b \right) {\hat \phi} {B_{y0}} - i{k_y}\rho_i^2{{\hat n}_i} {B_{y0}} = 0 ,
\end{equation}
\begin{equation}
i\omega {{\hat n}_e} - i{k_y}{{\hat u}_e} {B_{y0}} = 0 ,
\end{equation}
\begin{equation}
i\omega {\hat \psi}  - i\omega d_e^2{{\hat u}_e} - i{k_y} {\hat \phi}  {B_{y0}} + i{k_y}\rho_s^2{{\hat n}_e} {B_{y0}} = 0 ,
\end{equation}
with the closure relations:
\begin{equation}
k_\bot^2 {\hat \psi}  =  - {{\hat u}_e} + {\Gamma_0^{{1}/{2}}}\left( b \right) {{\hat u}_i} ,
\end{equation}
\begin{equation}
{\hat n}_e = {\Gamma_0^{{1}/{2}}}\left( b \right) {{\hat n}_i} + \frac{{\left( {{\Gamma_0}\left( b \right) - 1} \right)}}{{\rho_i^2}} {\hat \phi} .
\end{equation}
In Fourier space ${\Gamma _0}\left( b \right) = {e^{ - b}}{I_0}\left( b \right)$, where $b = k_\bot^2 \rho_i^2$ and $I_0$ is the modified Bessel function of the first kind. Hence, this system yields the following dispersion relation:
\begin{equation} \label{disp_rel}
\begin{array}{l}
\left[ {d_i^2\left( {1 + d_e^2k_ \bot ^2} \right) + d_e^2{\Gamma _0}\left( b \right)} \right]\left( {\dfrac{{{\omega ^4}}}{{k_y^4}}} \right) = \\
\quad \quad B_{y0}^2\left[ {\left( {d_i^2k_\bot ^2 + {\Gamma _0}\left( b \right)} \right)\rho_s^2 - \left( {1 + d_e^2k_ \bot ^2 + d_i^2k_ \bot ^2 - {\Gamma_0}\left( b \right)} \right)\dfrac{{\rho_i^2}}{{{\Gamma _0}\left( b \right) - 1}}} \right]\left( {\dfrac{{{\omega ^2}}}{{k_y^2}}} \right) + \\
\quad \quad  + \left( {\rho_s^2 + \rho _i^2} \right)B_{y0}^4 k_\bot ^2\dfrac{{\rho_i^2}}{{{\Gamma _0}\left( b \right) - 1}} .
\end{array}
\end{equation}
In the limit $d_i \to \infty$ this equation reduces to
\begin{equation} \label{disp_rel2}
\left( {1 + d_e^2 k_\bot^2} \right){\omega^2} = k_y^2 B_{y0}^2  k_\bot^2\left[ {\rho_s^2 - \frac{{\rho_i^2}}{{{\Gamma_0}\left( b \right) - 1}}} \right] ,
\end{equation}
which is the general dispersion relation of dispersive Alfv\'en waves in a homogeneous plasma. By assuming a regime such that $\beta_e \ll 2{{m_e}/{m_i}}$, we find the dispersion relation for the inertial Alfv\'en wave \cite{GB_1979}
\begin{equation}
\omega^2 = \frac{{k_y^2 B_{y0}^2}}{{1 + d_e^2k_\bot^2}} ,
\end{equation}
which reduces to the shear Alfv\'en wave in the limit $d_e^2 k_\bot^2 \ll 1$. In contrast, by assuming a regime such that $2{{m_e}/{m_i}} \ll \beta_e \ll 1$, we find the general dispersion relation for the kinetic Alfv\'en wave \cite{HC_1975}
\begin{equation}
\omega^2 = k_y^2 B_{y0}^2  k_\bot^2\left[ {\rho_s^2 - \frac{{\rho_i^2}}{{{\Gamma_0}\left( b \right) - 1}}} \right] . \label{KAW}
\end{equation}
In the limit $k_ \bot ^2\rho _i^2 \gg 1$, for which ${\Gamma_0}\left( b \right) \approx 0$, the above equation reduces to
\begin{equation}
\omega^2 = k_y^2 B_{y0}^2 k_\bot ^2\left( {\rho_s^2 + \rho_i^2} \right) . \label{KAW1}
\end{equation}
In the opposite limit $k_\bot ^2\rho _i^2 \ll 1$ we can expand the integral operator as ${\Gamma_0}\left( b \right) = 1 - b + \left( {{3}/{4}} \right)b^2 + {\mathcal{O}}\left( {{b^3}} \right)$, so that ${{b}/{{\left(1 - {\Gamma_0}\left( b \right)\right)}}} \approx {{1}/{{\left(1 - \left( {{3}/{4}} \right)b\right)}}} = 1 + \left( {{3}/{4}} \right)b + {\mathcal{O}}\left( {{b^2}} \right)$, and Eq. (\ref{KAW}) becomes	
\begin{equation}
{\omega ^2} = k_y^2 B_{y0}^2 \left[ {1 + k_ \bot ^2\rho _i^2\left( {\frac{3}{4} + \frac{{{T_e}}}{{{T_i}}}} \right)} \right] . \label{KAW2}
\end{equation}
On the right-hand side of the previous equation, the first term represents the shear Alfv\'en wave, whereas the other terms represent the finite Larmor radius corrections. Note that $ {\bf{k}} \cdot {{\bf{B}}_{eq}} = k_y B_{y0}$ since in this two-dimensional analysis $k_z = 0$. Therefore, the guide field $B_0$ enters only via the ion and ion-sound Larmor radius.


\clearpage

\begin{table}[h!]
\caption{\label{tab:table1} Maximum reconnection rate for different plasma parameters. The system size in the $y$-direction is $L_y=4\pi$.}
\begin{ruledtabular}
\begin{tabular}{ccccc}
$d_e(L)$ & $d_i(L)$ & $\rho_s(L)$ & $\rho_i(L)$ &$E_{z,X}^{\max}(v_{A,up} B_{y0,up})$\\
\hline
0.2    &  2     & 0.4   & 0.8  & 0.2827 \\
0.15   &  1.5   & 0.3   & 0.6  & 0.2076 \\
0.1    &  1     & 0.2   & 0.4  & 0.1418 \\
0.075  &  0.75  & 0.15  & 0.3  & 0.1249 \\
0.05   &  0.5   & 0.1   & 0.2  & 0.1176 \\
0.0375 &  0.375 & 0.075 & 0.15 & 0.1122 \\
0.025  &  0.25  & 0.05  & 0.1  & 0.1004 \\
\end{tabular}
\end{ruledtabular}
\end{table}

\end{document}